\documentclass[longauth]{aa}  

\usepackage{graphicx}
\usepackage{txfonts}
\usepackage[colorlinks=true,urlcolor=blue,linkcolor=blue,citecolor=blue]{hyperref}
\usepackage{threeparttable}
\usepackage{booktabs}
\usepackage{longtable}

\newcommand{\rearth}{R$_\oplus$}

\begin{document} 

   \title{Refining the properties of the TOI-178 system with\\ CHEOPS and TESS
   \thanks{The photometric data used in this work are only available at the CDS via anonymous ftp to \url{cdsarc.cds.unistra.fr} (\url{ftp://130.79.128.5}) or via \url{https://cdsarc.cds.unistra.fr/viz-bin/cat/J/A+A/}}\fnmsep
   \thanks{This study uses CHEOPS data observed as part of the Guaranteed Time Observation (GTO) programme CH\_PR100031.}}
   \titlerunning{A detailed study of the TOI-178 system}
   \authorrunning{L. Delrez et al.}

   \author{
L. Delrez\inst{1,2} $^{\href{https://orcid.org/0000-0001-6108-4808}{\includegraphics[scale=0.5]{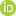}}}$,
A. Leleu\inst{3,4} $^{\href{https://orcid.org/0000-0003-2051-7974}{\includegraphics[scale=0.5]{orcid.jpg}}}$,
A. Brandeker\inst{5} $^{\href{https://orcid.org/0000-0002-7201-7536}{\includegraphics[scale=0.5]{orcid.jpg}}}$,
M. Gillon\inst{1} $^{\href{https://orcid.org/0000-0003-1462-7739}{\includegraphics[scale=0.5]{orcid.jpg}}}$,
M. J. Hooton\inst{6,4} $^{\href{https://orcid.org/0000-0003-0030-332X}{\includegraphics[scale=0.5]{orcid.jpg}}}$,
A. Collier Cameron\inst{7} $^{\href{https://orcid.org/0000-0002-8863-7828}{\includegraphics[scale=0.5]{orcid.jpg}}}$,
A. Deline\inst{3}, 
A. Fortier\inst{4,8} $^{\href{https://orcid.org/0000-0001-8450-3374}{\includegraphics[scale=0.5]{orcid.jpg}}}$,
D. Queloz\inst{9,6} $^{\href{https://orcid.org/0000-0002-3012-0316}{\includegraphics[scale=0.5]{orcid.jpg}}}$,
A. Bonfanti\inst{10} $^{\href{https://orcid.org/0000-0002-1916-5935}{\includegraphics[scale=0.5]{orcid.jpg}}}$,
V. Van Grootel\inst{2} $^{\href{https://orcid.org/0000-0003-2144-4316}{\includegraphics[scale=0.5]{orcid.jpg}}}$,
T. G. Wilson\inst{7} $^{\href{https://orcid.org/0000-0001-8749-1962}{\includegraphics[scale=0.5]{orcid.jpg}}}$,
J. A. Egger\inst{4} $^{\href{https://orcid.org/0000-0003-1628-4231}{\includegraphics[scale=0.5]{orcid.jpg}}}$,
Y. Alibert\inst{4} $^{\href{https://orcid.org/0000-0002-4644-8818}{\includegraphics[scale=0.5]{orcid.jpg}}}$,
R. Alonso\inst{11,12} $^{\href{https://orcid.org/0000-0001-8462-8126}{\includegraphics[scale=0.5]{orcid.jpg}}}$,
G. Anglada\inst{13,14} $^{\href{https://orcid.org/0000-0002-3645-5977}{\includegraphics[scale=0.5]{orcid.jpg}}}$,
J. Asquier\inst{15}, 
T. Bárczy\inst{16} $^{\href{https://orcid.org/0000-0002-7822-4413}{\includegraphics[scale=0.5]{orcid.jpg}}}$,
D. Barrado y Navascues\inst{17} $^{\href{https://orcid.org/0000-0002-5971-9242}{\includegraphics[scale=0.5]{orcid.jpg}}}$,
S. C. C. Barros\inst{18,19} $^{\href{https://orcid.org/0000-0003-2434-3625}{\includegraphics[scale=0.5]{orcid.jpg}}}$,
W. Baumjohann\inst{10} $^{\href{https://orcid.org/0000-0001-6271-0110}{\includegraphics[scale=0.5]{orcid.jpg}}}$,
M. Beck\inst{3} $^{\href{https://orcid.org/0000-0003-3926-0275}{\includegraphics[scale=0.5]{orcid.jpg}}}$,
T. Beck\inst{4}, 
W. Benz\inst{4,8} $^{\href{https://orcid.org/0000-0001-7896-6479}{\includegraphics[scale=0.5]{orcid.jpg}}}$,
N. Billot\inst{3} $^{\href{https://orcid.org/0000-0003-3429-3836}{\includegraphics[scale=0.5]{orcid.jpg}}}$,
X. Bonfils\inst{20} $^{\href{https://orcid.org/0000-0001-9003-8894}{\includegraphics[scale=0.5]{orcid.jpg}}}$,
L. Borsato\inst{21} $^{\href{https://orcid.org/0000-0003-0066-9268}{\includegraphics[scale=0.5]{orcid.jpg}}}$,\\
C. Broeg\inst{4,8} $^{\href{https://orcid.org/0000-0001-5132-2614}{\includegraphics[scale=0.5]{orcid.jpg}}}$,
M. Buder\inst{22}, 
J. Cabrera\inst{23} $^{\href{https://orcid.org/0000-0001-6653-5487}{\includegraphics[scale=0.5]{orcid.jpg}}}$,
V. Cessa\inst{4}, 
S. Charnoz\inst{24} $^{\href{https://orcid.org/0000-0002-7442-491X}{\includegraphics[scale=0.5]{orcid.jpg}}}$,
Sz. Csizmadia\inst{23} $^{\href{https://orcid.org/0000-0001-6803-9698}{\includegraphics[scale=0.5]{orcid.jpg}}}$,
P. E. Cubillos\inst{25,10} $^{\href{https://orcid.org/0000-0002-1347-2600}{\includegraphics[scale=0.5]{orcid.jpg}}}$,\\
M. B. Davies\inst{26} $^{\href{https://orcid.org/0000-0001-6080-1190}{\includegraphics[scale=0.5]{orcid.jpg}}}$,
M. Deleuil\inst{27} $^{\href{https://orcid.org/0000-0001-6036-0225}{\includegraphics[scale=0.5]{orcid.jpg}}}$,
O. D. S. Demangeon\inst{18,19} $^{\href{https://orcid.org/0000-0001-7918-0355}{\includegraphics[scale=0.5]{orcid.jpg}}}$,
B.-O. Demory\inst{8,4} $^{\href{https://orcid.org/0000-0002-9355-5165}{\includegraphics[scale=0.5]{orcid.jpg}}}$,
D. Ehrenreich\inst{3,28} $^{\href{https://orcid.org/0000-0001-9704-5405}{\includegraphics[scale=0.5]{orcid.jpg}}}$,\\
A. Erikson\inst{23}, 
L. Fossati\inst{10} $^{\href{https://orcid.org/0000-0003-4426-9530}{\includegraphics[scale=0.5]{orcid.jpg}}}$,
M. Fridlund\inst{29,30} $^{\href{https://orcid.org/0000-0002-0855-8426}{\includegraphics[scale=0.5]{orcid.jpg}}}$,
D. Gandolfi\inst{31} $^{\href{https://orcid.org/0000-0001-8627-9628}{\includegraphics[scale=0.5]{orcid.jpg}}}$,
M. Güdel\inst{32}, 
J. Hasiba\inst{10}, 
S. Hoyer\inst{27} $^{\href{https://orcid.org/0000-0003-3477-2466}{\includegraphics[scale=0.5]{orcid.jpg}}}$,\\
K. G. Isaak\inst{33} $^{\href{https://orcid.org/0000-0001-8585-1717}{\includegraphics[scale=0.5]{orcid.jpg}}}$,
J. M. Jenkins\inst{34} $^{\href{https://orcid.org/0000-0002-4715-9460}{\includegraphics[scale=0.5]{orcid.jpg}}}$,
L. L. Kiss\inst{35,36}, 
J. Laskar\inst{37} $^{\href{https://orcid.org/0000-0003-2634-789X}{\includegraphics[scale=0.5]{orcid.jpg}}}$,
D. W. Latham\inst{38} $^{\href{https://orcid.org/0000-0001-9911-7388}{\includegraphics[scale=0.5]{orcid.jpg}}}$,
A. Lecavelier des Etangs\inst{39} $^{\href{https://orcid.org/0000-0002-5637-5253}{\includegraphics[scale=0.5]{orcid.jpg}}}$,\\
M. Lendl\inst{3} $^{\href{https://orcid.org/0000-0001-9699-1459}{\includegraphics[scale=0.5]{orcid.jpg}}}$,
C. Lovis\inst{3} $^{\href{https://orcid.org/0000-0001-7120-5837}{\includegraphics[scale=0.5]{orcid.jpg}}}$,
R. Luque\inst{40} $^{\href{https://orcid.org/0000-0002-4671-2957}{\includegraphics[scale=0.5]{orcid.jpg}}}$,
D. Magrin\inst{21} $^{\href{https://orcid.org/0000-0003-0312-313X}{\includegraphics[scale=0.5]{orcid.jpg}}}$,
P. F. L. Maxted\inst{41} $^{\href{https://orcid.org/0000-0003-3794-1317}{\includegraphics[scale=0.5]{orcid.jpg}}}$,
C. Mordasini\inst{4,8} $^{\href{https://orcid.org/0000-0002-1013-2811}{\includegraphics[scale=0.5]{orcid.jpg}}}$,
V. Nascimbeni\inst{21} $^{\href{https://orcid.org/0000-0001-9770-1214}{\includegraphics[scale=0.5]{orcid.jpg}}}$,
G. Olofsson\inst{5} $^{\href{https://orcid.org/0000-0003-3747-7120}{\includegraphics[scale=0.5]{orcid.jpg}}}$,
R. Ottensamer\inst{42}, 
I. Pagano\inst{43} $^{\href{https://orcid.org/0000-0001-9573-4928}{\includegraphics[scale=0.5]{orcid.jpg}}}$,
E. Pallé\inst{11} $^{\href{https://orcid.org/0000-0003-0987-1593}{\includegraphics[scale=0.5]{orcid.jpg}}}$,
G. Peter\inst{22} $^{\href{https://orcid.org/0000-0001-6101-2513}{\includegraphics[scale=0.5]{orcid.jpg}}}$,
G. Piotto\inst{21,44} $^{\href{https://orcid.org/0000-0002-9937-6387}{\includegraphics[scale=0.5]{orcid.jpg}}}$,
D. Pollacco\inst{45} $^{\href{https://orcid.org/0000-0001-9850-9697}{\includegraphics[scale=0.5]{orcid.jpg}}}$,\\
R. Ragazzoni\inst{21,44} $^{\href{https://orcid.org/0000-0002-7697-5555}{\includegraphics[scale=0.5]{orcid.jpg}}}$,
N. Rando\inst{15}, 
H. Rauer\inst{23,46,47} $^{\href{https://orcid.org/0000-0002-6510-1828}{\includegraphics[scale=0.5]{orcid.jpg}}}$,
I. Ribas\inst{13,14} $^{\href{https://orcid.org/0000-0002-6689-0312}{\includegraphics[scale=0.5]{orcid.jpg}}}$,
G. Ricker\inst{48,49} $^{\href{https://orcid.org/0000-0003-2058-6662}{\includegraphics[scale=0.5]{orcid.jpg}}}$,
N. C. Santos\inst{18,19} $^{\href{https://orcid.org/0000-0003-4422-2919}{\includegraphics[scale=0.5]{orcid.jpg}}}$,\\
G. Scandariato\inst{43} $^{\href{https://orcid.org/0000-0003-2029-0626}{\includegraphics[scale=0.5]{orcid.jpg}}}$,
S. Seager\inst{50,48,49,51}, 
D. Ségransan\inst{3} $^{\href{https://orcid.org/0000-0003-2355-8034}{\includegraphics[scale=0.5]{orcid.jpg}}}$,
A. E. Simon\inst{4} $^{\href{https://orcid.org/0000-0001-9773-2600}{\includegraphics[scale=0.5]{orcid.jpg}}}$,
A. M. S. Smith\inst{23} $^{\href{https://orcid.org/0000-0002-2386-4341}{\includegraphics[scale=0.5]{orcid.jpg}}}$,
S. G. Sousa\inst{18} $^{\href{https://orcid.org/0000-0001-9047-2965}{\includegraphics[scale=0.5]{orcid.jpg}}}$,\\
M. Steller\inst{10} $^{\href{https://orcid.org/0000-0003-2459-6155}{\includegraphics[scale=0.5]{orcid.jpg}}}$,
Gy. M. Szabó\inst{52,53}, 
N. Thomas\inst{4}, 
S. Udry\inst{3} $^{\href{https://orcid.org/0000-0001-7576-6236}{\includegraphics[scale=0.5]{orcid.jpg}}}$,
R. Vanderspek\inst{48} $^{\href{https://orcid.org/0000-0001-6763-6562}{\includegraphics[scale=0.5]{orcid.jpg}}}$,
J. Venturini\inst{3} $^{\href{https://orcid.org/0000-0001-9527-2903}{\includegraphics[scale=0.5]{orcid.jpg}}}$,
V. Viotto\inst{21} $^{\href{https://orcid.org/0000-0001-5700-9565}{\includegraphics[scale=0.5]{orcid.jpg}}}$,\\
N. A. Walton\inst{54} $^{\href{https://orcid.org/0000-0003-3983-8778}{\includegraphics[scale=0.5]{orcid.jpg}}}$,
J. N. Winn\inst{55}
          }

   \institute{
\label{inst:1} Astrobiology Research Unit, Université de Liège, Allée du 6 Août 19C, B-4000 Liège, Belgium \and
\label{inst:2} Space sciences, Technologies and Astrophysics Research (STAR) Institute, Université de Liège, Allée du 6 Août 19C, 4000 Liège, Belgium \and
\label{inst:3} Observatoire Astronomique de l'Université de Genève, Chemin Pegasi 51, CH-1290 Versoix, Switzerland \and
\label{inst:4} Physikalisches Institut, University of Bern, Sidlerstrasse 5, 3012 Bern, Switzerland \and
\label{inst:5} Department of Astronomy, Stockholm University, AlbaNova University Center, 10691 Stockholm, Sweden \and
\label{inst:6} Cavendish Laboratory, JJ Thomson Avenue, Cambridge CB3 0HE, UK \and
\label{inst:7} Centre for Exoplanet Science, SUPA School of Physics and Astronomy, University of St Andrews, North Haugh, St Andrews KY16 9SS, UK \and
\label{inst:8} Center for Space and Habitability, University of Bern, Gesellschaftsstrasse 6, 3012 Bern, Switzerland \and
\label{inst:9} ETH Zurich, Department of Physics, Wolfgang-Pauli-Strasse 2, CH-8093 Zurich, Switzerland \and
\label{inst:10} Space Research Institute, Austrian Academy of Sciences, Schmiedlstrasse 6, A-8042 Graz, Austria \and
\label{inst:11} Instituto de Astrofisica de Canarias, 38200 La Laguna, Tenerife, Spain \and
\label{inst:12} Departamento de Astrofisica, Universidad de La Laguna, 38206 La Laguna, Tenerife, Spain \and
\label{inst:13} Institut de Ciencies de l'Espai (ICE, CSIC), Campus UAB, Can Magrans s/n, 08193 Bellaterra, Spain \and
\label{inst:14} Institut d'Estudis Espacials de Catalunya (IEEC), 08034 Barcelona, Spain \and
\label{inst:15} ESTEC, European Space Agency, 2201AZ, Noordwijk, NL \and
\label{inst:16} Admatis, 5. Kandó Kálmán Street, 3534 Miskolc, Hungary \and
\label{inst:17} Depto. de Astrofisica, Centro de Astrobiologia (CSIC-INTA), ESAC campus, 28692 Villanueva de la Cañada (Madrid), Spain \and
\label{inst:18} Instituto de Astrofisica e Ciencias do Espaco, Universidade do Porto, CAUP, Rua das Estrelas, 4150-762 Porto, Portugal \and
\label{inst:19} Departamento de Fisica e Astronomia, Faculdade de Ciencias, Universidade do Porto, Rua do Campo Alegre, 4169-007 Porto, Portugal \and
\label{inst:20} Université Grenoble Alpes, CNRS, IPAG, 38000 Grenoble, France \and
\label{inst:21} INAF, Osservatorio Astronomico di Padova, Vicolo dell'Osservatorio 5, 35122 Padova, Italy \and
\label{inst:22} Institute of Optical Sensor Systems, German Aerospace Center (DLR), Rutherfordstrasse 2, 12489 Berlin, Germany \and
\label{inst:23} Institute of Planetary Research, German Aerospace Center (DLR), Rutherfordstrasse 2, 12489 Berlin, Germany \and
\label{inst:24} Université de Paris, Institut de physique du globe de Paris, CNRS, F-75005 Paris, France \and
\label{inst:25} INAF, Osservatorio Astrofisico di Torino, Via Osservatorio, 20, I-10025 Pino Torinese To, Italy \and
\label{inst:26} Centre for Mathematical Sciences, Lund University, Box 118, 221 00 Lund, Sweden \and
\label{inst:27} Aix Marseille Univ, CNRS, CNES, LAM, 38 rue Frédéric Joliot-Curie, 13388 Marseille, France \and
\label{inst:28} Centre Vie dans l'Univers, Faculté des sciences, Université de Genève, Quai Ernest-Ansermet 30, CH-1211 Genève 4, Switzerland \and
\label{inst:29} Leiden Observatory, University of Leiden, PO Box 9513, 2300 RA Leiden, The Netherlands \and
\label{inst:30} Department of Space, Earth and Environment, Chalmers University of Technology, Onsala Space Observatory, 439 92 Onsala, Sweden \and
\label{inst:31} Dipartimento di Fisica, Universita degli Studi di Torino, via Pietro Giuria 1, I-10125, Torino, Italy \and
\label{inst:32} University of Vienna, Department of Astrophysics, Türkenschanzstrasse 17, 1180 Vienna, Austria \and
\label{inst:33} Science and Operations Department - Science Division (SCI-SC), Directorate of Science, European Space Agency (ESA), European Space Research and Technology Centre (ESTEC),
Keplerlaan 1, 2201-AZ Noordwijk, The Netherlands \and
\label{inst:34} NASA Ames Research Center, Moffett Field, CA 94035, USA \and
\label{inst:35} Konkoly Observatory, Research Centre for Astronomy and Earth Sciences, 1121 Budapest, Konkoly Thege Miklós út 15-17, Hungary \and
\label{inst:36} ELTE E\"otv\"os Lor\'and University, Institute of Physics, P\'azm\'any P\'eter s\'et\'any 1/A, 1117 Budapest, Hungary \and
\label{inst:37} IMCCE, UMR8028 CNRS, Observatoire de Paris, PSL Univ., Sorbonne Univ., 77 av. Denfert-Rochereau, 75014 Paris, France \and
\label{inst:38} Center for Astrophysics | Harvard \& Smithsonian, 60 Garden Street, Cambridge, MA, 02138, USA \and
\label{inst:39} Institut d'astrophysique de Paris, UMR7095 CNRS, Université Pierre \& Marie Curie, 98bis blvd. Arago, 75014 Paris, France \and
\label{inst:40} Department of Astronomy and Astrophysics, University of Chicago, Chicago, IL 60637, USA \and
\label{inst:41} Astrophysics Group, Keele University, Staffordshire, ST5 5BG, United Kingdom \and
\label{inst:42} Department of Astrophysics, University of Vienna, Tuerkenschanzstrasse 17, 1180 Vienna, Austria \and
\label{inst:43} INAF, Osservatorio Astrofisico di Catania, Via S. Sofia 78, 95123 Catania, Italy \and
\label{inst:44} Dipartimento di Fisica e Astronomia "Galileo Galilei", Universita degli Studi di Padova, Vicolo dell'Osservatorio 3, 35122 Padova, Italy \and
\label{inst:45} Department of Physics, University of Warwick, Gibbet Hill Road, Coventry CV4 7AL, United Kingdom \and
\label{inst:46} Zentrum für Astronomie und Astrophysik, Technische Universität Berlin, Hardenbergstr. 36, D-10623 Berlin, Germany \and
\label{inst:47} Institut für Geologische Wissenschaften, Freie Universität Berlin, 12249 Berlin, Germany \and
\label{inst:48} Kavli Institute for Astrophysics and Space Research, Massachusetts Institute of Technology, Cambridge, MA 02139, USA \and
\label{inst:49} Department of Physics, Massachusetts Institute of Technology, Cambridge, MA 02139, USA \and
\label{inst:50} Department of Earth, Atmospheric and Planetary Science, Massachusetts Institute of Technology, 77 Massachusetts Avenue, Cambridge, MA 02139, USA \and
\label{inst:51} Department of Aeronautics and Astronautics, MIT, 77 Massachusetts Avenue, Cambridge, MA 02139, USA \and
\label{inst:52} ELTE E\"otv\"os Lor\'and University, Gothard Astrophysical Observatory, 9700 Szombathely, Szent Imre h. u. 112, Hungary \and
\label{inst:53} MTA-ELTE Exoplanet Research Group, 9700 Szombathely, Szent Imre h. u. 112, Hungary \and
\label{inst:54} Institute of Astronomy, University of Cambridge, Madingley Road, Cambridge, CB3 0HA, United Kingdom \and
\label{inst:55} Department of Astrophysical Sciences, Princeton University, 4 Ivy Lane, Princeton, NJ 08544, USA 
             }

   \date{Received ...; accepted ...}

 
  \abstract
   {The TOI-178 system consists of a nearby late K-dwarf transited by six planets in the super-Earth to mini-Neptune regime, with radii ranging from $\sim$1.1 to 2.9 $R_{\oplus}$ and orbital periods between 1.9 and 20.7 days. All planets but the innermost one form a chain of Laplace resonances. Mass estimates derived from a preliminary radial velocity (RV) dataset suggest that the planetary densities do not decrease in a monotonic way with the orbital distance to the star, contrary to what one would expect based on simple formation and evolution models.}
   {To improve the characterisation of this key system and prepare for future studies (in particular with JWST), we perform a detailed photometric study based on 40 new CHEOPS visits, one new TESS sector, as well as previously published CHEOPS, TESS, and NGTS data.}
   {First we update the parameters of the host star using the new parallax from \textit{Gaia} EDR3. We then perform a global analysis of the 100 transits contained in our data to refine the physical and orbital parameters of the six planets and study their transit timing variations (TTVs). We also use our extensive dataset to place constraints on the radii and orbital periods of potential additional transiting planets in the system.}
   {Our analysis significantly refines the transit parameters of the six planets, most notably their radii, for which we now obtain relative precisions $\lesssim$3\%, with the exception of the smallest planet $b$ for which the precision is 5.1\%. Combined with the RV mass estimates, the measured TTVs allow us to constrain the eccentricities of planets $c$ to $g$, which are found to be all below 0.02, as expected from stability requirements. Taken alone, the TTVs also suggest a higher mass for planet $d$ than the one estimated from the RVs, which had been found to yield a surprisingly low density for this planet. However, the masses derived from the current TTV dataset are very prior-dependent and further observations, over a longer temporal baseline, are needed to deepen our understanding of this iconic planetary system.}
  {}


   \maketitle
%
\section{Introduction}
\label{sec:intro}

Studying the relation between the internal composition of planets found in multi-planet systems and their architecture (i.e. orbital properties) is crucial to improve our understanding of the formation and evolution of planetary systems. In this context, planetary systems forming chains of three-body Laplace resonances, where each consecutive pair of planets are in (or close to) a mean-motion resonance (MMR), are of particular interest. Indeed, the fine-tuning and fragility of such orbital configurations ensure that no significant scattering or collision event has taken place since the formation of the planets in the protoplanetary disc \citep[e.g.][]{2016Natur.533..509M}. Hence, these systems are real goldmines for constraining the outcome of protoplanetary discs and provide important anchors for planet formation models. 

To date, chains of Laplace resonances have only been observed for a few systems: GJ 876 \citep{2010ApJ...719..890R}, Kepler-60 \citep{2016MNRAS.455L.104G}, Kepler-80 \citep{2016AJ....152..105M}, Kepler-223 \citep{2016Natur.533..509M}, TRAPPIST-1 \citep{2017Natur.542..456G,2017NatAs...1E.129L}, K2-138 \citep{2018AJ....155...57C,2019A&A...631A..90L}, TOI-178 (\citealt{Leleu2021}, hereafter L21), and \hbox{TOI-1136} \citep{2023AJ....165...33D}. All these systems, except GJ 876, are transiting, which provides an opportunity to constrain the masses and eccentricities of the planets via their transit timing variations (TTVs), which may have detectable amplitudes thanks to the proximity of each pair of planets to a MMR. For stars that are bright enough, it is also possible to obtain radial velocity (RV) measurements, that can provide complementary constraints on the planetary masses and orbital parameters. Out of the transiting systems cited above, only K2-138, TOI-1136, and TOI-178 have published RV measurements so far, the other ones being too faint in the visible ($V$-mag$\gtrsim$14). In this work, we focus on the latter of these three systems.

The nearby ($\sim$63 pc) late K-type star TOI-178 was initially flagged by the \textit{Transiting Exoplanet Survey Satellite} (TESS, \citealt{2015JATIS...1a4003R}) as a potential host to three transiting sub-Neptunes with orbital periods of 6.56, 9.96, and 10.35 days, based on data from its Sector 2. The 0.4-day difference in the orbital periods of the two outer planetary candidates led \cite{2019A&A...624A..46L} to hypothesise that they occupied a horseshoe orbital configuration. Thanks to an intensive photometric follow-up with the \textit{CHaracterising ExOPlanets Satellite} (CHEOPS, \citealt{2021ExA....51..109B}), L21 demonstrated that this was actually not the case, and revealed instead a compact system of at least six transiting planets in the super-Earth to mini-Neptune regime, with radii ranging from $\sim$1.1 to 2.9 $R_{\oplus}$ and orbital periods of 1.91, 3.24, 6.56, 9.96, 15.23, and 20.71 days. The five outer planets form a 2:4:6:9:12 chain of Laplace resonances, while the innermost planet $b$ lies just outside the 3:5 MMR with planet $c$, which could indicate that it was previously part of the chain but was then pulled away, possibly by tidal forces.

Using RV measurements obtained with the Echelle SPectrograph for Rocky Exoplanets and Stable Spectroscopic Observations (ESPRESSO, \citealt{2021A&A...645A..96P}) installed at ESO's Very Large Telescope, L21 were also able to derive preliminary estimates for the masses of the planets, and thus their bulk densities (when combined with the radii inferred from the transit photometry). The planetary densities that they found show important variations from planet to planet, jumping for example from $\sim$1 to 0.2 $\rho_{\oplus}$ between planets $c$ and $d$. By doing a Bayesian internal structure analysis, they showed that the two innermost planets are likely to be mostly rocky, which could indicate that they have lost their primordial gas envelope through atmospheric escape, while all the other planets appear to contain significant amounts of water and/or gas (see also the independent internal structure analysis by \citealt{2022A&A...660A.102A}). Interestingly, it seems that the amount of gas in the planets does not vary as a monotonic function of the orbital distance to the star, as opposed to what one would expect from simple formation and evolution models and unlike other known systems in a chain of Laplace resonances. The most notable outlier is planet $d$, which seems to have a larger gas mass than planet $e$ (with a probability of 92\%), although the latter is more massive and at a larger distance from the star. This is surprising for two reasons. First, from a formation perspective, one would expect that the mass of the primordial gas envelope is a growing function of the total planetary mass. Second, from an evolution point of view, one would also expect that atmospheric evaporation is more effective for planets that are closer to the star. Based on these considerations, we would thus expect planet $d$ to have a smaller gas mass than planet $e$. Another possible outlier is planet $f$, which appears to be the most massive planet in the system, but may still have less gas than planet $e$ (with a probability of $\sim$60\%) despite being located further away from the star.

However, the planetary densities on which these results are based are rather poorly constrained (precision $\gtrsim$30\%). In particular, the planetary masses presented by L21 were derived using only 46 RV data points, a very limited dataset for such a complex system. Further RV observations are thus needed to confirm and refine these preliminary mass estimates. Complementary constraints on the masses and eccentricities of the planets could also be obtained by monitoring their TTVs, which are expected to be measurable for all but the innermost planet, with predicted amplitudes ranging from a few minutes for the inner planets to a few tens of minutes for the outer ones (L21). Such transit follow-up observations would also be useful to refine the transit parameters of the planets and their radii. Improving the overall characterisation of the TOI-178 system is essential to optimally prepare the atmospheric follow-up observations that are scheduled on JWST/NIRSpec (PI: M. Hooton) for three of its planets ($b$, $d$, and $g$) and support the interpretation of the resulting transmission spectra.

These considerations motivated the work presented here, which consists in a detailed photometric study of the TOI-178 system, based on 40 new CHEOPS visits, one new TESS sector, as well as previously published data. Our extensive dataset contains 100 transits of the six planets in total, about twice more than the transit dataset presented in L21. The paper is structured as follows. In \hbox{Sect. \ref{sec:star}}, we update the properties of the host star using the new parallax from \textit{Gaia} EDR3. Sect. \ref{sec:obs} describes all the observations that we used in our work, with a particular focus on the new data presented in this paper. In Sect. \ref{sec:analysis_top_section}, we present our detailed analysis of all these data, including a global transit analysis to refine the system parameters and measure the individual transit timings (Sect. \ref{sec:analysis}), as well as a search for possible additional transiting planets in the data and an assessment of their detection limits (Sect. \ref{sec:detection_limits}). In Sect. \ref{sec:dynamical_analysis}, we present a dynamical analysis of the individual transit timings measured for the five outer planets, before concluding in Sect. \ref{sec:conclusion}.

\section{Stellar properties}
\label{sec:star}

L21 already provided a thorough characterisation of the host star. Table \ref{tab:star} gives the effective temperature ($T_{\mathrm{eff}}$), surface gravity \hbox{(log $g_\star$)}, metallicity ([Fe/H]), and projected rotational velocity ($v$ sin $i_\star$) that they derived from a detailed spectroscopic analysis of the 46 ESPRESSO high-resolution spectra. We refine here the stellar radius of TOI-178 in a similar fashion as in L21, but using updated \textit{Gaia} EDR3 photometry and parallax values. In brief, we employed a Markov-Chain Monte Carlo (MCMC) modified infrared flux method (IRFM; \citealt{Blackwell1977}; \citealt{Schanche2020}) to compute the bolometric flux by fitting \textit{Gaia} \citep{GaiaCollaboration2021}, 2MASS \citep{Skrutskie2006}, and WISE \citep{Wright2010} broadband photometry with stellar atmospheric models \citep{Castelli2003}. In this process, the spectroscopic parameters from L21 were used as priors on stellar atmospheric model selection. The bolometric flux was then converted into stellar effective temperature and angular diameter, which was subsequently used to determine the stellar radius ($R_{\star}$) of TOI-178 using the offset-corrected \textit{Gaia} EDR3 parallax \citep{Lindegren2021}. Via this method, we obtained $R_{\star}$ = $0.662 \pm 0.010$ $R_\odot$, which is similar to the value reported in L21.

\begin{table}[hbt!]
\caption{Updated properties of the host star TOI-178.}
\begin{tabular}{l c c}
\toprule
\toprule
\textbf{Property (unit)} & \textbf{Value} & \textbf{Source} \\
\midrule
\multicolumn{2}{l}{\textit{Astrometric properties}} & \\
RA (J2000) & 00:29:12.49 & [1] \\
Dec (J2000) & $-$30:27:14.86 & [1] \\
$\mathrm{\mu_{RA}}$ (mas $\mathrm{yr}^{-1}$) & $150.032 \pm 0.028$ & [1] \\
$\mathrm{\mu_{Dec}}$ (mas $\mathrm{yr}^{-1}$) & $-87.132 \pm 0.030$ & [1] \\
Parallax (mas) & $15.900 \pm 0.031$ & [1] \\
Distance (pc) & $62.89 \pm 0.12$ & from parallax \\
\midrule
\multicolumn{2}{l}{\textit{Photometric magnitudes}} & \\
$G$ (mag) & $11.1575 \pm 0.0028$ & [1]\\
$G_{\rm BP}$ (mag) & $11.8398 \pm 0.0029$ & [1]\\
$G_{\rm RP}$ (mag) & $10.3602 \pm 0.0038$ & [1]\\
$J$ (mag) & $9.372 \pm 0.021$ & [2] \\
$H$ (mag) & $8.761 \pm 0.023$ & [2] \\
$K$ (mag) & $8.656 \pm 0.021$ & [2] \\
$W$1 (mag) & $8.573 \pm 0.022$ & [3]\\
$W$2 (mag) & $8.64 \pm 0.02$ & [3]\\
\midrule
\multicolumn{2}{l}{\textit{Spectroscopic and derived properties}} & \\
$T_{\mathrm{eff}}$ (K) & $4316 \pm 70$ & Spectroscopy [4] \\ 
log $g_\star$ (cgs) & $4.45 \pm 0.15$ & Spectroscopy [4] \\ 
$\mathrm{[Fe/H]}$ (dex) & $-0.23 \pm 0.05$ & Spectroscopy [4] \\
$v$ sin $i_\star$ (km $\mathrm{s}^{-1}$) & $1.5 \pm 0.3$ & Spectroscopy [4] \\
$R_\star$ ($R_\odot$) & $0.662 \pm 0.010$ & IRFM [5] \\
$M_\star$ ($M_\odot$) & $0.647_{-0.029}^{+0.030}$ & Isochrones [5] \\ 
$t_\star$ (Gyr) & $6.0_{-5.0}^{+6.8}$ & Isochrones [5] \\
$L_\star$ ($L_\odot$) & $0.136\pm0.010$ & from $R_\star$ and $T_{\mathrm{eff}}$ [5] \\
$\rho_\star$ ($\rho_\odot$) & $2.23\pm0.14$ & from $R_\star$ and $M_\star$ [5] \\ 
\bottomrule
\bottomrule
\end{tabular}
\textbf{References.} [1] \textit{Gaia} EDR3 \citep{GaiaCollaboration2021}; [2] 2MASS \citep{Skrutskie2006}; [3] WISE \citep{Wright2010}; [4] \cite{Leleu2021}; [5] this work (see Sect. \ref{sec:star}).
\label{tab:star}
\end{table}

Taking advantage of the $R_{\star}$ revision, we re-derived the isochronal mass ($M_{\star}$) and age ($t_{\star}$) following the same procedure outlined in L21. We inputted [Fe/H], $T_{\mathrm{eff}}$, and $R_{\star}$ into two different stellar evolutionary models, namely PARSEC\footnote{\textit{PA}dova and T\textit{R}ieste \textit{S}tellar \textit{E}volutionary \textit{C}ode: \url{http://stev.oapd.inaf.it/cgi-bin/cmd}} v1.2S \citep{marigo17} and CLES \citep[Code Liégeois d'Évolution Stellaire,][]{scuflaire08} to obtain two pairs of mass and age estimates. In particular, the first pair ($M_{\star,1}$, $t_{\star,1}$) was computed by the isochrone placement algorithm \citep{bonfanti15,bonfanti16}, which interpolates the provided input values within pre-computed grids of PARSEC isochrones and tracks. The routine convergence was further aided by the $v\sin{i_\star}$-based gyrochronological relation (with $v\sin{i_\star}$ the projected rotational velocity) implemented within the isochrone placement, as described in \citet{bonfanti16}. The second pair ($M_{\star,2}$, $t_{\star,2}$), instead, was retrieved by CLES, which generates the best-fit `on-the-fly' stellar track, following the Levenberg-Marquardt minimisation scheme presented in \citet{salmon21}. After carefully checking the mutual consistency of the two respective pairs of outcomes through the $\chi^2$-based criterion broadly discussed in \citet{bonfanti21}, we finally merged the two results and obtained $M_{\star}=0.647_{-0.029}^{+0.030}\,M_{\odot}$ and $t_{\star}=6.0_{-5.0}^{+6.8}$ Gyr. Those values are similar to those reported in L21. Our revised stellar parameters are presented at the bottom of Table \ref{tab:star}.

\section{Data}
\label{sec:obs}

In this section, we describe all the photometric data that we used in our work. Table \ref{tab:transit_summary} summarises the number of transits obtained for each planet and facility, with a total of 100 transits observed for the system. This is about twice more than the transit dataset presented in L21.

\begin{table}[hbt!]
\caption{Number of transits observed for each planet and facility.}
\centering
\begin{tabular}{lcccccc}
\toprule
\toprule
         &  \multicolumn{6}{c}{Planet} \\
Facility &  b & c & d & e & f & g \\
\midrule
\vspace{0.1cm}
CHEOPS & 19 & 9 & 5 & 4 & 4 & 4 \\
\vspace{0.1cm}
TESS &  23 & 14 & 6 & 4 & 4 & 2 \\
\vspace{0.1cm}
NGTS & 1 & -- & -- & -- & -- & 1 \\
\midrule
\vspace{0.1cm}
Total number of transits & 43 & 23 & 11 & 8 & 8 & 7 \\
\bottomrule
\bottomrule
\end{tabular}
\label{tab:transit_summary}
\end{table}

\subsection{CHEOPS}
\label{sec:cheops_data}

We obtained 44 visits (observation runs) of TOI-178 with CHEOPS \citep{2021ExA....51..109B} in total, of which only four were presented previously in L21. These observations were acquired between 21 July 2020 and 18 October 2021 as part of the Guaranteed Time Observations (GTO) program and are summarised in Table \ref{tab:CHEOPS_visits}. Since CHEOPS revolves around the Earth on a low-altitude ($\sim$700 km) Sun-synchronous orbit, the data show some interruptions corresponding to occultations of the target by the Earth or passages through the South Atlantic Anomaly (SAA). For our TOI-178 visits, the resulting observing efficiencies (fraction of time used for science observations) vary between 45 and 93\% depending on the date of observation. Due to the relative faintness of TOI-178 ($G$-mag=11.15) for CHEOPS, we used the maximum exposure time of 60 seconds for all visits. 

As part of our CHEOPS dataset is the near-continuous \hbox{11-day} observation performed in August 2020, split into two visits for scheduling reasons, that was previously presented in L21. Among the other 42 visits, 23 of them were scheduled to cover transits of the known planets, while the remaining 19 were `fillers', i.e. observations that are carried out when CHEOPS has no time-constrained or higher-priority observations. In the case of \hbox{TOI-178}, the goal of these fillers was to search for other potential transiting planets in the system but they did not reveal any transit-like signal (see Sect. \ref{sec:detection_limits}). Together, the CHEOPS data covered 19, 9, 5, 4, 4, and 4 transits of TOI-178\,b, c, d, e, f, and g, respectively.

The raw data of each visit were automatically processed with the CHEOPS Data Reduction Pipeline (DRP version 13.1.0; \citealt{2020A&A...635A..24H}). In short, the DRP calibrates the raw images (event flagging, bias and gain corrections, linearisation, dark current, and flat field corrections), corrects them for environmental effects (cosmic rays, background, and smearing trails from nearby stars), and performs aperture photometry to extract the target's flux for four different apertures. Using the \texttt{pycheops} package\footnote{\url{https://github.com/pmaxted/pycheops}} \citep{2022MNRAS.514...77M} to analyse the different light curves, we found that the best precision is obtained in this case with the default photometric aperture (25 pixels). Owing to the extended and irregular shape of the CHEOPS Point Spread Function (PSF) and the fact that the field rotates around the target along the spacecraft’s nadir-locked orbit \citep{2021ExA....51..109B}, nearby background stars can introduce time-variable flux contamination in the photometric aperture, in phase with the spacecraft roll angle (see e.g. \citealt{2020A&A...643A..94L, bonfanti21, 2022MNRAS.514...77M}). The DRP also provides an estimation of this contamination by using \textit{Gaia} DR2 catalogue \citep{2018A&A...616A...1G} and a PSF template to simulate CHEOPS images of the field of view. For our TOI-178 observations, this contamination varies between 0.03 and 0.09\% of the target's flux and is mostly modulated by the rotation around the target of a nearby background star with $G$-mag=13.3 at a projected sky distance of 60.8\arcsec. The light curves were corrected for this contamination.

To get an independent photometric extraction, we also reduced the data with PIPE\footnote{\url{https://github.com/alphapsa/PIPE}} (Brandeker et al. in prep., \citealt{2021A&A...651L..12M}, \citealt{2021A&A...654A.159S}, \citealt{2022A&A...659A..74D}, \citealt{2022A&A...659L...4B}), a PSF photometry package developed specifically for CHEOPS that has demonstrated an improved precision for faint stars ($G$-mag $\gtrsim$11) such as TOI-178 in a previous work \citep{2021A&A...651L..12M}. PIPE first uses a principal component analysis (PCA) approach to derive a PSF template library from the data series. The first five principal components (PCs) together with a constant background are then used to fit the individual PSFs of each image using a least-squares minimisation and measure the target's flux. The number of PCs to use is a trade-off between following systematic PSF changes and overfitting the noise. For faint stars such as TOI-178, the mean PSF (first PC) is sufficient for a good extraction, and attempts to model the PSF better with more PCs usually introduce noise in the extracted light curve. Some advantages of using PSF photometry rather than aperture photometry for faint targets are that: (1) the contributions to the signal of each pixel over the PSF are weighted according to noise so that higher S/N photometry can be extracted; (2) cosmic rays and bad pixels (both hot and telegraphic) are easier to filter out or give lower weight in the fitting process; (3) PSF photometry is less sensitive to contamination from nearby background stars; (4) the background is fit simultaneously with the PSF for the same pixels, which can be an advantage if there is some spatial structure. 

For each visit, we estimated the photometric precision by computing the median absolute deviation (MAD) of the difference between two consecutive data points (d$f = f_{i+1}-f_i$) of the light curve. This metric is robust to outliers and removes correlated signals (e.g. transits). Table \ref{tab:CHEOPS_visits} gives the MAD that we obtained for both the DRP and PIPE. We found a significant improvement for PIPE, of 18\% on average in terms of MAD. We thus decided to use the PIPE light curves in our global analysis.

\subsection{TESS}

TESS \citep{2015JATIS...1a4003R} observed TOI-178 for the first time during Cycle 1/Sector 2 of its primary mission (22 August -- 20 September 2018). These data, obtained with a two-minute cadence, were previously presented in L21 and we include them in our global analysis. TESS observed again TOI-178 during Cycle 3/Sector 29 of its extended mission, from 26 August to 22 September 2020. The observations were acquired on CCD 3 of camera 2. The data were processed with the TESS Science Processing Operations Center (SPOC) pipeline \citep{2016SPIE.9913E..3EJ} at NASA Ames Research Center. We retrieved the 2-minute cadence Presearch Data Conditioning Simple Aperture Photometry (PDCSAP, \citealt{2012PASP..124..985S,2012PASP..124.1000S,2014PASP..126..100S}) from the Mikulski Archive for Space Telescopes\footnote{\url{https://archive.stsci.edu}} (MAST), using the default quality bitmask. Together, the two TESS sectors covered 23, 14, 6, 4, 4, and 2 transits of TOI-178\,b, c, d, e, f, and g, respectively. 

\subsection{NGTS}

We also included in our global analysis the light curves obtained with the Next Generation Transit Survey (NGTS, \citealt{2018MNRAS.475.4476W}) that were previously published in L21: one transit of \hbox{planet $b$} observed simultaneously with six telescopes on 11 September 2019 and one transit of planet $g$ observed on 12 October 2019 using seven telescopes, thus a total of 13 light curves. We refer the reader to L21 and references therein for more information about these NGTS data and their reduction.


\begin{figure*}[hbt!]
    \centering
    \includegraphics[width=0.49\textwidth]{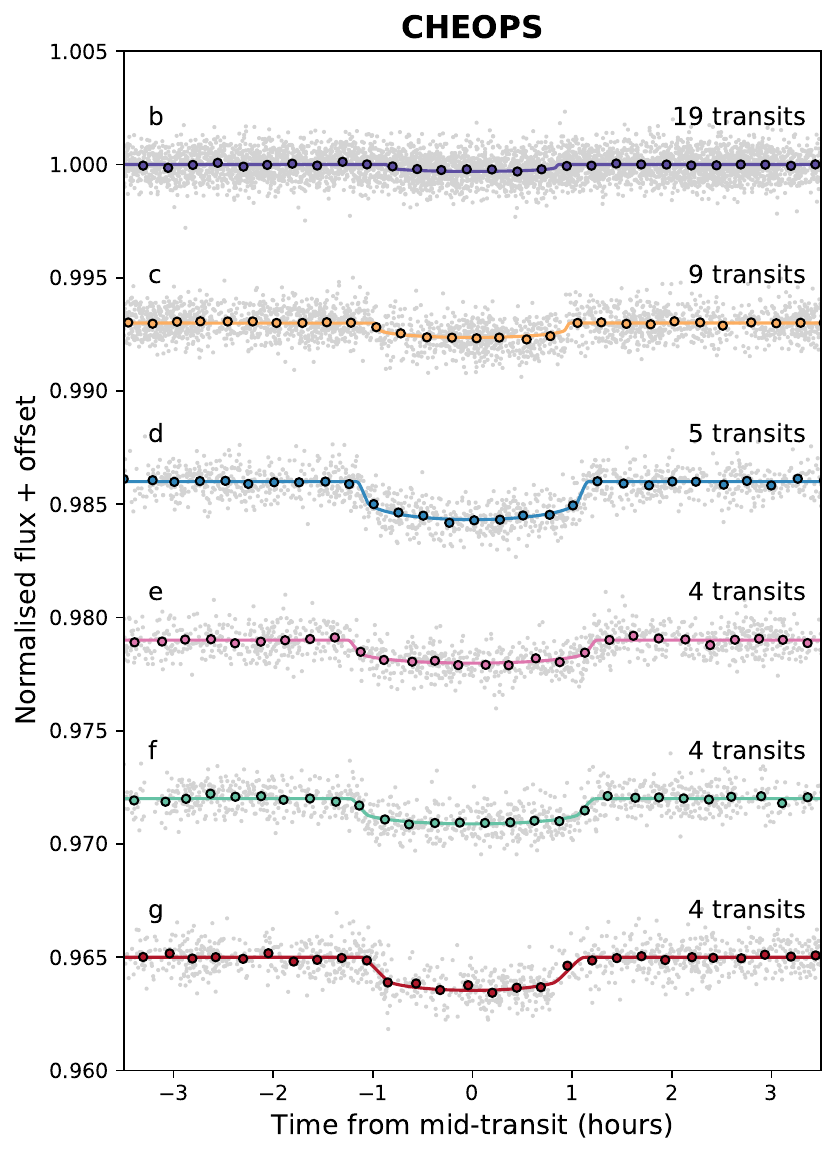}
    \includegraphics[width=0.49\textwidth]{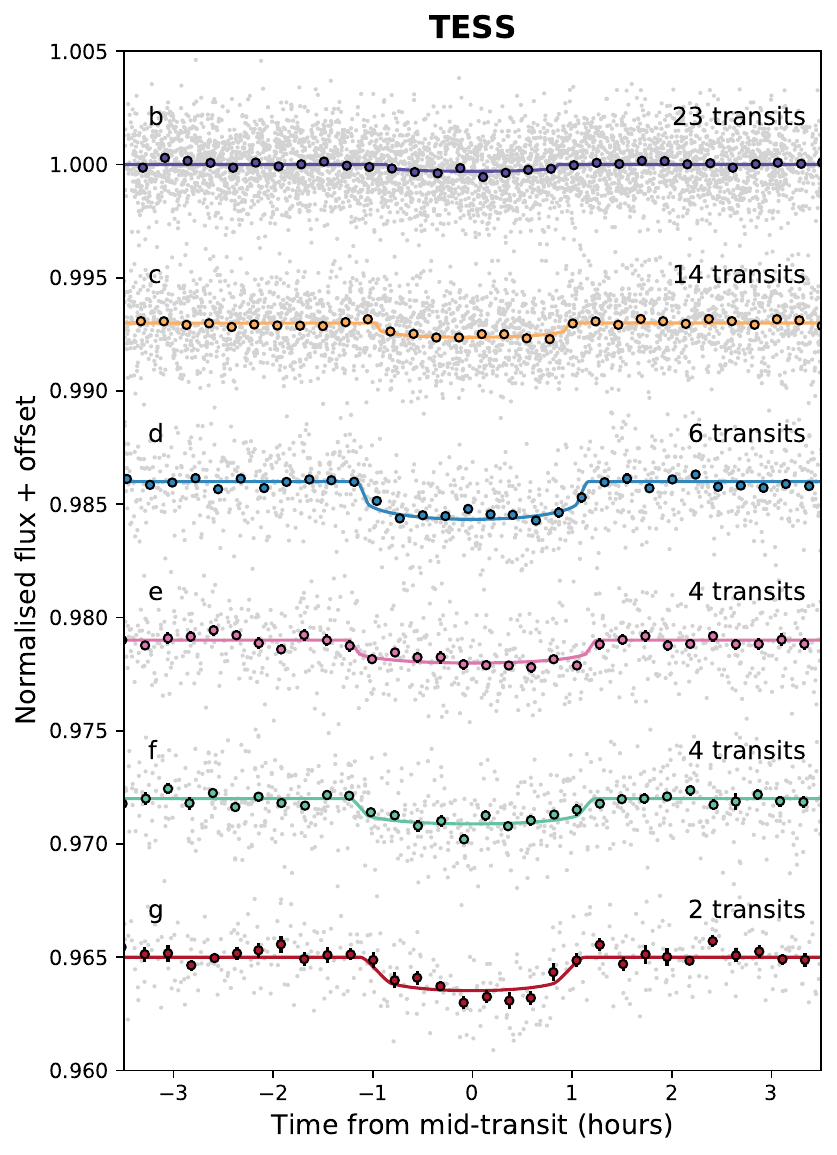}
    \caption{Phase-folded detrended transit photometry of the TOI-178 planets obtained with CHEOPS (left) and TESS (right). For each planet, the photometry was corrected for the measured TTVs and for the transit signals of the other planets. The unbinned data points are shown in grey, while the coloured circles with error bars correspond to 15-min bins. The coloured lines show the best-fit transit models. The number of transits combined in each phase-folded light curve is also indicated in the plot.}
    \label{fig:folded_LCs}
\end{figure*}

\begin{figure*}
    \centering
    \includegraphics[width=0.49\textwidth]{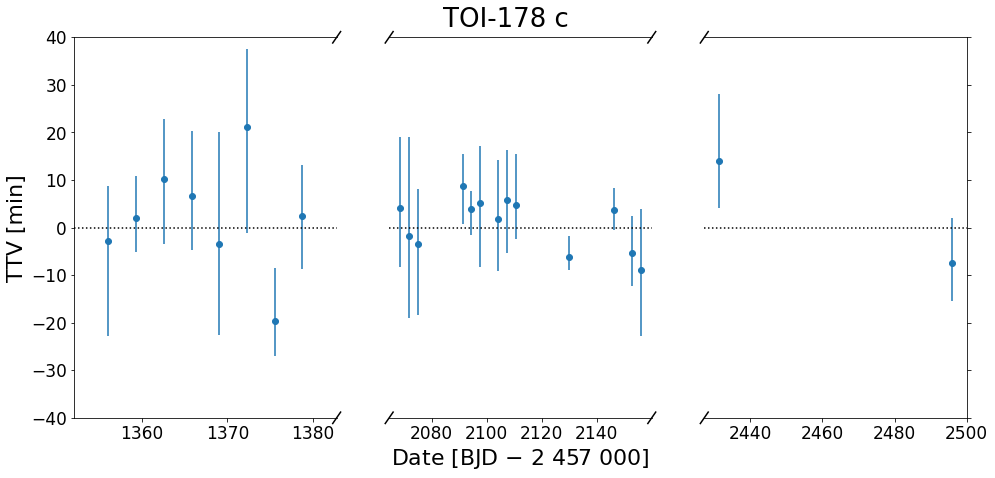}
    \includegraphics[width=0.49\textwidth]{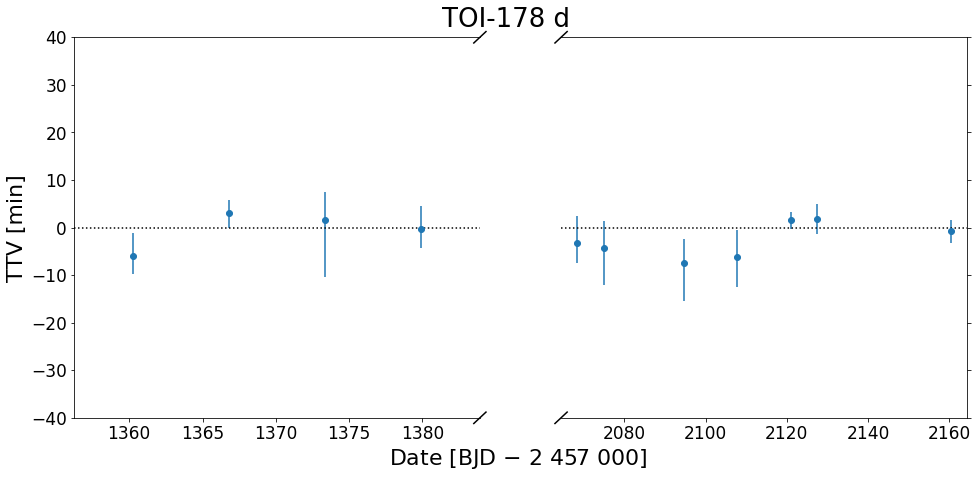}
    \includegraphics[width=0.49\textwidth]{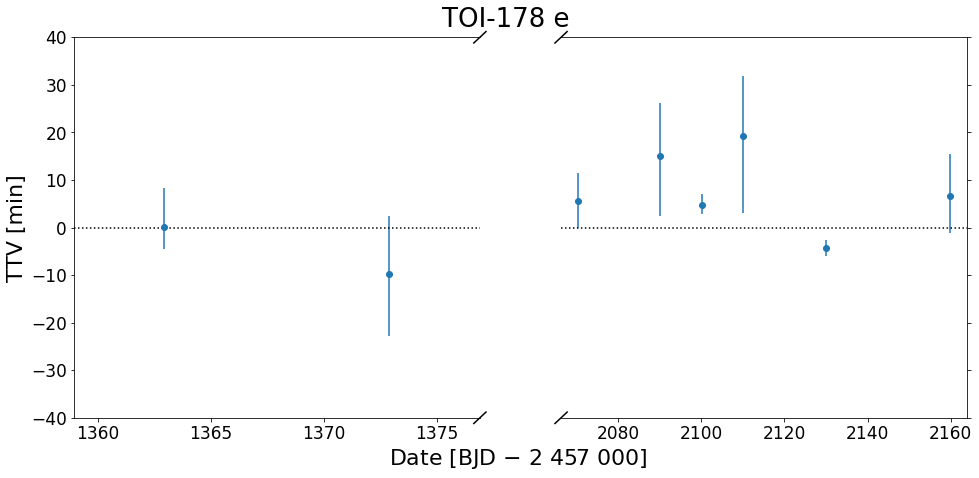}
    \includegraphics[width=0.49\textwidth]{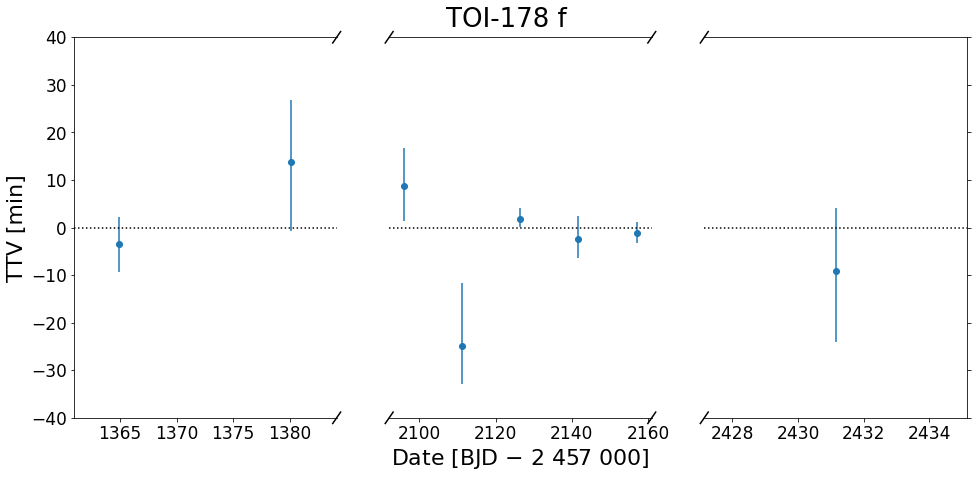}
    \includegraphics[width=0.49\textwidth]{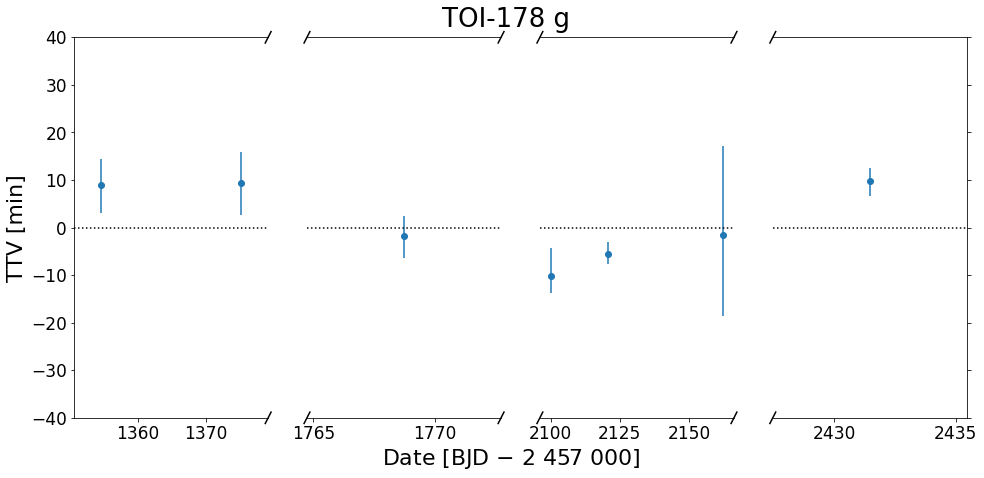}
    \caption{TTVs (in minutes) deduced from our global transit analysis (see Sect. \ref{sec:analysis}) for the five outer TOI-178 planets. These TTVs are relative to the updated mean transit ephemerides given in Tables \ref{tab:transit_parameters1} (planets $b$, $c$, and $d$) and \ref{tab:transit_parameters2} (planets $e$, $f$, and $g$). For each planet, we zoomed around the epochs of the observed transits in the $x$-axes for a better visual assessment of the measured TTVs.}
    \label{fig:ttv_zoom}
\end{figure*}

\section{Data analysis}
\label{sec:analysis_top_section}

\subsection{Global transit analysis}
\label{sec:analysis}

We performed a joint fit of all the transit photometry described in Sect. \ref{sec:obs} using the most recent version of the adaptive MCMC algorithm presented in \citeauthor{2012A&A...542A...4G} (\citeyear{2012A&A...542A...4G}, see also \citealt{2014A&A...563A..21G}). In order to reduce processing times in this global transit analysis, we only used the portions of the data which contain transits (leaving enough out-of-transit data for proper modelling of the photometric baseline, see below), thus ending up with a total of 96 light curves (summarised in Table \ref{tab:baselines}). These light curves were modelled using the quadratic limb-darkening transit model of \cite{2002ApJ...580L.171M} multiplied by a photometric baseline model, different for each light curve, aimed at representing the photometric variations caused by other astrophysical, instrumental, or environmental effects. For each light curve, we explored a large range of baseline models, including polynomials, cubic splines, or Gaussian Processes with respect to e.g. time, background, target's location on the detector, spacecraft roll angle, telescope tube temperature (for CHEOPS), airmass (for NGTS), or combinations of these parameters. Table \ref{tab:baselines} gives the baseline models selected for each light curve based on the Bayesian Information Criterion (BIC, \citealt{1978AnSta...6..461S}). The minimal baseline model is a simple constant to account for any out-of-transit flux offset.

The model parameters sampled by the MCMC were:
\begin{itemize}
    \item for each planet, the transit depth ($\mathrm{d}F=R_{\mathrm{p}}^2/R_\star^2$ where $R_{\rm{p}}$ is the radius of the planet and $R_\star$ is the stellar radius) and the cosine of the orbital inclination (\hbox{cos $i_{\rm{p}}$});
    \item the log of the stellar density (log $\rho_\star$), the log of the stellar mass (log $M_\star$), and the effective temperature ($T_{\rm{eff}}$);
    \item for the five outer planets, the transit timing variation (TTV, in minutes) of each transit with respect to the transit ephemerides defined by the orbital period ($P$) and the mid-transit time ($T_0$) reported in Tables 3 and 4 of L21;
    \item the log of the orbital period \hbox{(log $P$)} and the mid-transit time ($T_0$) of TOI-178\,b (we assumed a linear transit ephemeris for this planet as it is not part of the Laplace resonant chain and is thus not expected to show any significant TTVs, see L21);
    \item for each bandpass (CHEOPS, TESS, and NGTS), the combinations $q_1=(u_1+u_2)^2$ and $q_2=0.5\,u_1 (u_1+u_2)^{-1}$ of the quadratic limb-darkening coefficients ($u_1$ and $u_2$), following the triangular sampling scheme advocated by \cite{2013MNRAS.435.2152K}.
\end{itemize}
We note that our model thus did not include dynamical interactions between the components of the TOI-178 system; it assumed that there are no mutual interactions between the planets (fixed orbital periods) and then measured TTVs with respect to that Keplerian model. We also assumed circular orbits for all the planets (as justified in Sect. \ref{sec:dynamical_analysis}, see also L21) and thus set their respective $\sqrt{e}$ cos $\omega$ and \hbox{$\sqrt{e}$ sin $\omega$} values (with $e$ the eccentricity and $\omega$ the argument of periastron) to zero. Given the large number of light curves, the coefficients of the photometric baseline models were not sampled by the MCMC but determined by a least-squares fit to the residuals at each step of the procedure. This approach allowed us to avoid a dramatic increase in the number of sampled parameters while still fitting the correlated noise simultaneously with the transits (instead of pre-detrending the data), thus ensuring a better propagation of the uncertainties to the derived system parameters of interest (see also similar TTV studies of the \hbox{TRAPPIST-1} planets by \citealt{2018MNRAS.475.3577D} and \citealt{2020A&A...640A.112D}).

The prior distributions used in our analysis are summarised in Table \ref{tab:priors}. We assumed normal priors for $M_\star$, $R_\star$, and $T_{\rm{eff}}$ based on the values and uncertainties derived in Sect. \ref{sec:star} (Table \ref{tab:star}). We also computed normal priors for the quadratic limb-darkening coefficients ($u_1$ and $u_2$) in each bandpass using the \texttt{LDCU}\footnote{\url{https://github.com/delinea/LDCU}} code \citep{2022A&A...659A..74D}, which builds on the method described by \cite{2015MNRAS.450.1879E} to generate limb-darkening coefficients from two libraries of synthetic stellar spectra, ATLAS \citep{kurucz_atlas_1979} and PHOENIX \citep{husser_phoenix_2013}, while propagating the uncertainties on the stellar parameters and models. 

We first ran a preliminary MCMC chain of 50 000 steps to estimate the two scaling factors, $\beta_{w}$ and $\beta_{r}$ (Table \ref{tab:baselines}), to be applied to the photometric error bars of each light curve to account respectively for over- or under-estimated white noise and the presence of residual correlated (red) noise (for details, see \citealt{2012A&A...542A...4G} and references therein). With the corrected photometric error bars, we then ran two chains of 250\,000 steps each (including 25\% burn-in) and checked their convergence by using the statistical test of \cite{1992StaSc...7..457G}, ensuring that the test values for all sampled parameters were <1.01.

\begin{table*}[!htbp]
    \centering
    \begin{tabular}{lcc}
    \hline\hline
    \vspace{0.1cm}
    \textbf{Parameter (unit)} & \textbf{Leleu et al.\ (2021)} & \textbf{This work}  \\
    \hline
    \multicolumn{3}{l}{\textbf{TOI-178 b}} \\
    \vspace{0.1cm}
    Transit depth, d$F$ (ppm)                          & $263_{-30}^{+32}$                & $249_{-23}^{+24}$ \\
    \vspace{0.1cm}
    Transit impact parameter, $b$ ($R_\star$)                     & $0.17_{-0.13}^{+0.19}$           & $0.25_{-0.13}^{+0.15}$           \\
    \vspace{0.1cm}
    Orbital period, $P$ (d)                             & $1.914558 \pm 0.000018$          & $1.914554_{-0.000015}^{+0.000018}$ \\
    \vspace{0.1cm}
    Mid-transit time, $T_0$ ($\mathrm{BJD_{TDB}} - 2\:450\:000$) & $8\,741.6365_{-0.0030}^{+0.0043}$ & $8\,931.1779_{-0.0017}^{+0.0023}$  \\
    \vspace{0.1cm}
    Transit duration, $W$ (hours)                         & $1.692_{-0.086}^{+0.056}$        & $1.705_{-0.076}^{+0.033}$ \\
    \vspace{0.1cm}
    Orbital inclination, $i_{\rm{p}}$ (deg)                        & $88.8_{-1.3}^{+0.8}$             & $88.33_{-1.03}^{+0.88}$ \\
    \vspace{0.1cm}
    Orbital semi-major axis, $a_{\rm{p}}$ (au)                            & $0.02607 \pm 0.00078$            & $0.02609_{-0.00034}^{+0.00031}$ \\
    \vspace{0.1cm}
    Scale parameter, $a_{\mathrm{p}}/R_{\star}$           & $8.61_{-0.22}^{+0.21}$ & $8.47_{-0.13}^{+0.12}$ \\
    \vspace{0.1cm}
    Radius, $R_p$ (\rearth)                     & $1.152_{-0.070}^{+0.073}$ (6.3\%) & $1.142\pm0.058$ (5.1\%)             \\
    \vspace{0.1cm}
    Stellar irradiation, $S_{\mathrm{p}}$ ($S_{\oplus}$)     & -- & $198 \pm 14$ \\
    \hline
    \multicolumn{3}{l}{\textbf{TOI-178 c}} \\
    \vspace{0.1cm}
    Transit depth, d$F$ (ppm)                          & $551_{-59}^{+68}$                & $542_{-24}^{+29}$ \\
    \vspace{0.1cm}
    Transit impact parameter, $b$ ($R_\star$)                     & $0.34_{-0.23}^{+0.30}$           & $0.24 \pm 0.14$           \\
    \vspace{0.1cm}
    Orbital period, $P$ (d)                             & $3.238450_{-0.000019}^{+0.000020}$ & $3.238449 \pm 0.000011$ \\
    \vspace{0.1cm}
    Mid-transit time, $T_0$ ($\mathrm{BJD_{TDB}} - 2\:450\:000$) & $8\,741.4783_{-0.0029}^{+0.0034}$ & $8\,926.0680 \pm 0.0011$  \\
    \vspace{0.1cm}
    Transit duration, $W$ (hours)                         & $1.95_{-0.25}^{+0.15}$           & $2.045_{-0.083}^{+0.047}$ \\
    \vspace{0.1cm}
    Orbital inclination, $i_{\rm{p}}$ (deg)                        & $88.4_{-1.6}^{+1.1}$             & $88.86_{-0.70}^{+0.69}$ \\
    \vspace{0.1cm}
    Orbital semi-major axis, $a_{\rm{p}}$ (au)                            & $0.0370 \pm 0.0011$              & $0.03703_{-0.00048}^{+0.00044}$ \\
    \vspace{0.1cm}
    Scale parameter, $a_{\mathrm{p}}/R_{\star}$            & $12.23_{-0.31}^{+0.29}$ & $12.03_{-0.19}^{+0.18}$ \\
    \vspace{0.1cm}
    Radius, $R_p$ (\rearth)                     & $1.669_{-0.099}^{+0.114}$ (6.8\%) & $1.685_{-0.051}^{+0.052}$ (3.1\%) \\
    \vspace{0.1cm}
    Stellar irradiation, $S_{\mathrm{p}}$ ($S_{\oplus}$)     & -- & $98.3 \pm 7.2$ \\
    \vspace{0.1cm}
    Orbital eccentricity (TTVs), $e_{\rm{p}}$ & -- & $0.0073_{-0.0051}^{+0.0083}$ \\
    \hline
    \multicolumn{3}{l}{\textbf{TOI-178 d}} \\
    \vspace{0.1cm}
    Transit depth, d$F$ (ppm)                          & $1313_{-65}^{+64}$                & $1414_{-51}^{+50}$ \\
    \vspace{0.1cm}
    Transit impact parameter, $b$ ($R_\star$)                     & $0.485_{-0.060}^{+0.051}$         & $0.530_{-0.035}^{+0.032}$           \\
    \vspace{0.1cm}
    Orbital period, $P$ (d)                             & $6.557700 \pm 0.000016$          &  $6.557721 \pm 0.000013$ \\
    \vspace{0.1cm}
    Mid-transit time, $T_0$ ($\mathrm{BJD_{TDB}} - 2\:450\:000$) & $8\,747.14623_{-0.00095}^{+0.00087}$ &  $8\,760.26373 \pm 0.00074$ \\
    \vspace{0.1cm}
    Transit duration, $W$ (hours)                         & $2.346_{-0.046}^{+0.047}$        & $2.323_{-0.034}^{+0.041}$ \\
    \vspace{0.1cm}
    Orbital inclination, $i_{\rm{p}}$ (deg)                        & $88.58_{-0.18}^{+0.20}$          & $88.42_{-0.11}^{+0.12}$ \\
    \vspace{0.1cm}
    Orbital semi-major axis, $a_{\rm{p}}$ (au)                            & $0.0592 \pm 0.0018$              & $0.05927_{-0.00077}^{+0.00070}$ \\
    \vspace{0.1cm}
    Scale parameter, $a_{\mathrm{p}}/R_{\star}$            & $19.57_{-0.49}^{+0.47}$ & $19.25_{-0.30}^{+0.28}$ \\
    \vspace{0.1cm}
    Radius, $R_p$ (\rearth)                     & $2.572_{-0.078}^{+0.075}$ (3.0\%) & $2.717_{-0.061}^{+0.066}$ (2.4\%) \\
    \vspace{0.1cm}
    Stellar irradiation, $S_{\mathrm{p}}$ ($S_{\oplus}$)    & -- & $38.4 \pm 2.8$ \\
    \vspace{0.1cm}
    Orbital eccentricity (TTVs), $e_{\rm{p}}$ & -- & $0.010_{-0.007}^{+0.011}$ \\
    \hline\hline
    \end{tabular}
    \caption{Properties of the TOI-178\,b, c, and d planets based on our global transit analysis (see Sect. \ref{sec:analysis}). For planets $c$ and $d$, we also give the eccentricities derived from our TTV analysis (see Sect. \ref{sec:dynamical_analysis}). For each parameter, we indicate the median of the posterior distribution, along with the 1-$\sigma$ credible intervals. For the planetary radii, we also give the relative uncertainties in brackets.}
    \label{tab:transit_parameters1}
\end{table*}

\begin{table*}[!htbp]
    \centering
    \begin{tabular}{lcc}
    \hline\hline
    \vspace{0.1cm}
    \textbf{Parameter (unit)} & \textbf{Leleu et al.\ (2021)} & \textbf{This work}  \\
    \hline
    \multicolumn{3}{l}{\textbf{TOI-178 e}} \\
    \vspace{0.1cm}
    Transit depth, d$F$ (ppm)                          & $968_{-71}^{+69}$                & $917_{-36}^{+33}$ \\
    \vspace{0.1cm}
    Transit impact parameter, $b$ ($R_\star$)                     & $0.583_{-0.066}^{+0.046}$        & $0.595_{-0.025}^{+0.024}$           \\
    \vspace{0.1cm}
    Orbital period, $P$ (d)                             & $9.961881 \pm 0.000042$          & $9.961815 \pm 0.000090$\\
    \vspace{0.1cm}
    Mid-transit time, $T_0$ ($\mathrm{BJD_{TDB}} - 2\:450\:000$) & $8\,751.4658_{-0.0019}^{+0.0016}$ & $8\,761.4267 \pm 0.0032$  \\
    \vspace{0.1cm}
    Transit duration, $W$ (hours)                         & $2.501_{-0.077}^{+0.106}$        & $2.517_{-0.041}^{+0.045}$ \\
    \vspace{0.1cm}
    Orbital inclination, $i_{\rm{p}}$ (deg)                        & $88.71_{-0.13}^{+0.16}$          & $88.662_{-0.071}^{+0.066}$ \\
    \vspace{0.1cm}
    Orbital semi-major axis, $a_{\rm{p}}$ (au)                            & $0.07833_{-0.00103}^{+0.00093}$   & $0.02609_{-0.00034}^{+0.00031}$ \\
    \vspace{0.1cm}
    Scale parameter, $a_{\mathrm{p}}/R_{\star}$          & $25.87_{-0.65}^{+0.62}$ & $25.44_{-0.40}^{+0.37}$ \\
    \vspace{0.1cm}
    Radius, $R_p$ (\rearth)                     & $2.207_{-0.090}^{+0.088}$ (4.1\%) & $2.189_{-0.058}^{+0.053}$ (2.6\%)             \\
    \vspace{0.1cm}
    Stellar irradiation, $S_{\mathrm{p}}$ ($S_{\oplus}$)     & -- & $22.0 \pm 1.6$ \\
    \vspace{0.1cm}
    Orbital eccentricity (TTVs), $e_{\rm{p}}$ & -- & $0.0080_{-0.0057}^{+0.0100}$ \\
    \hline
    \multicolumn{3}{l}{\textbf{TOI-178 f}} \\
    \vspace{0.1cm}
    Transit depth, d$F$ (ppm)                          & $1037_{-90}^{+94}$                & $1154_{-53}^{+42}$ \\
    \vspace{0.1cm}
    Transit impact parameter, $b$ ($R_\star$)                     & $0.765_{-0.031}^{+0.027}$          & $0.753_{-0.020}^{+0.015}$        \\
    \vspace{0.1cm}
    Orbital period, $P$ (d)                             & $15.231915_{-0.000095}^{+0.000105}$ & $15.231951 \pm 0.000095$ \\
    \vspace{0.1cm}
    Mid-transit time, $T_0$ ($\mathrm{BJD_{TDB}} - 2\:450\:000$) & $8\,745.7178_{-0.0027}^{+0.0023}$ & $8\,898.0349 \pm 0.0017$  \\
    \vspace{0.1cm}
    Transit duration, $W$ (hours)                         & $2.348_{-0.087}^{+0.097}$           & $2.446_{-0.043}^{+0.052}$ \\
    \vspace{0.1cm}
    Orbital inclination, $i_{\rm{p}}$ (deg)                        & $88.723_{-0.069}^{+0.071}$        & $88.723_{-0.044}^{+0.047}$ \\
    \vspace{0.1cm}
    Orbital semi-major axis, $a_{\rm{p}}$ (au)                            & $0.1039 \pm 0.0031$              & $0.1040_{-0.0014}^{+0.0012}$ \\
    \vspace{0.1cm}
    Scale parameter, $a_{\mathrm{p}}/R_{\star}$          & $34.33_{-0.87}^{+0.82}$ & $33.76_{-0.53}^{+0.49}$ \\
    \vspace{0.1cm}
    Radius, $R_p$ (\rearth)                     & $2.287_{-0.110}^{+0.108}$ (4.8\%) & $2.455_{-0.073}^{+0.061}$ (3.0\%) \\ 
    \vspace{0.1cm}
    Stellar irradiation, $S_{\mathrm{p}}$ ($S_{\oplus}$)     & --  & $12.47 \pm 0.91$ \\
    \vspace{0.1cm}
    Orbital eccentricity (TTVs), $e_{\rm{p}}$ & -- & $0.0105_{-0.0061}^{+0.0071}$ \\
    \hline
    \multicolumn{3}{l}{\textbf{TOI-178 g}} \\
    \vspace{0.1cm}
    Transit depth, d$F$ (ppm)                          & $1633_{-139}^{+157}$                & $1620_{-62}^{+54}$ \\
    \vspace{0.1cm}
    Transit impact parameter, $b$ ($R_\star$)                     & $0.866_{-0.019}^{+0.017}$         & $0.863_{-0.0105}^{+0.0097}$           \\
    \vspace{0.1cm}
    Orbital period, $P$ (d)                             & $20.70950_{-0.00011}^{+0.00014}$  & $20.70991 \pm 0.00015 $ \\
    \vspace{0.1cm}
    Mid-transit time, $T_0$ ($\mathrm{BJD_{TDB}} - 2\:450\:000$) & $8\,748.0302_{-0.0017}^{+0.0023}$ & $8\,893.0016 \pm 0.0026$ \\
    \vspace{0.1cm}
    Transit duration, $W$ (hours)                         & $2.167_{-0.082}^{+0.090}$        & $2.218_{-0.058}^{+0.061}$ \\
    \vspace{0.1cm}
    Orbital inclination, $i_{\rm{p}}$ (deg)                        & $88.823_{-0.047}^{+0.045}$          & $88.806_{-0.024}^{+0.023}$ \\
    \vspace{0.1cm}
    Orbital semi-major axis, $a_{\rm{p}}$ (au)                            & $0.1275_{-0.0039}^{+0.0038}$     & $0.1276_{-0.0017}^{+0.0015}$ \\
    \vspace{0.1cm}
    Scale parameter, $a_{\mathrm{p}}/R_{\star}$          & $42.13_{-1.06}^{+1.01}$ & $41.43_{-0.64}^{+0.61}$ \\
    \vspace{0.1cm}
    Radius, $R_p$ (\rearth)                     & $2.87_{-0.13}^{+0.14}$ (4.9\%) & $2.908_{-0.070}^{+0.068}$ (2.4\%) \\
    \vspace{0.1cm}
    Stellar irradiation, $S_{\mathrm{p}}$ ($S_{\oplus}$)     & -- & $8.28_{-0.61}^{+0.60}$ \\
    \vspace{0.1cm}
    Orbital eccentricity (TTVs), $e_{\rm{p}}$ & -- & $0.0056_{-0.0039}^{+0.0058}$ \\
    \hline\hline
    \end{tabular}
    \caption{Properties of the TOI-178\,e, f, and g planets based on our global transit analysis (see Sect. \ref{sec:analysis}). We also give the eccentricities derived from our TTV analysis (see Sect. \ref{sec:dynamical_analysis}). For each parameter, we indicate the median of the posterior distribution, along with the 1-$\sigma$ credible intervals. For the planetary radii, we also give the relative uncertainties in brackets.}
    \label{tab:transit_parameters2}
\end{table*}

Fig.~\ref{fig:folded_LCs} shows for each planet the phase-folded (TTV-corrected) detrended transit photometry from CHEOPS (left) and TESS (right), with the corresponding best-fit transit models. The medians and 1-$\sigma$ credible intervals of the posterior distributions obtained for the system parameters are given in Tables \ref{tab:transit_parameters1} (planets $b$, $c$, and $d$) and \ref{tab:transit_parameters2} (planets $e$, $f$, and $g$). The transit parameters of the six planets are significantly refined compared to L21, most notably their radii, for which we now obtain a relative precision $\lesssim$3\% for all planets, except the smallest planet $b$ for which the precision is 5.1\%. Table \ref{tab:timings} presents the individual transit timings that we obtained for the five outer planets. For each of these planets, we performed a linear fit of these transit timings as a function of their epochs to derive updated mean transit ephemerides, which are also given in Tables \ref{tab:transit_parameters1} and \ref{tab:transit_parameters2}. The reduced $\chi^2$ values of these linear fits are 0.8, 0.9, 2.9, 1.4, and 4.3 for TOI-178\,c, d, e, f, and g, respectively. The TTVs with respect to the updated ephemerides are given in Table \ref{tab:timings} and shown in Fig.~\ref{fig:ttv_zoom}. Only the latest transit of TOI-178\,g observed with CHEOPS shows a TTV different from zero at the $\sim$3-$\sigma$ level. A detailed dynamical analysis of the measured transit timings will be presented in Sect.~\ref{sec:dynamical_analysis}.

\subsection{Search for additional transiting planets and detection limits}
\label{sec:detection_limits}

In this section, we first aim to use our large photometric dataset to search for additional transiting planets in the system. We then perform transit injection-and-recovery tests to assess the detection limits of the data and thus place constraints on the radii and orbital periods of potential additional transiting planets in the system.

\subsubsection{Search for additional transiting planets}

\noindent 
We first searched the TESS data for additional transit signals using the \texttt{SHERLOCK}\footnote{The \texttt{SHERLOCK} (Searching for Hints of Exoplanets fRom Lightcurves Of spaCe-based seeKers) code is fully available on GitHub: \url{https://github.com/franpoz/SHERLOCK}} pipeline presented in \cite{2020A&A...641A..23P, 2023A&A...672A..70P}. \texttt{SHERLOCK} downloads the PDCSAP light curve from MAST and, using the \texttt{Wōtan} package \citep{2019AJ....158..143H}, applies a biweight sliding filter with varying window sizes to detrend the data. The motivation behind this multi-detrend approach is related to the risk of removing transit signals when detrending the light curve, especially short and shallow ones. Each detrended light curve and the original PDCSAP light curve are then searched for transit signals using the \texttt{transit least squares} (TLS) algorithm \citep{2019A&A...623A..39H}. The transit search is carried out in a loop: once a signal is found, it is stored and masked, and then the search keeps running until no more signals above a user-defined signal detection efficiency (SDE, \citealt{2019A&A...623A..39H}) threshold are found in the dataset. Each of these search-find-mask iterations is called a `run'. Here, we analysed the 2 TESS sectors simultaneously and tested 10 different window sizes between 0.2 and 1.2 days for the detrending. We searched the PDCSAP and the 10 detrended light curves for periodic signals with orbital periods ranging from 0.5 to 60 days. For each run, we selected the signal which was found in the greatest number of light curves among these 11 light curves, and with the highest SDE. We recovered the 6 known planets in the first 6 runs in the following order: first TOI-178\,d, then TOI-178\,c, TOI-178\,e, TOI-178\,f, TOI-178\,b, and finally TOI-178\,g. In the subsequent runs, we did not find any other promising signal with a SDE$\geq$5 that could hint at the presence of extra transiting planets in the system.

We then ran \texttt{SHERLOCK} on the CHEOPS data following the same procedure. For this purpose, we pre-detrended the CHEOPS photometry using a cubic spline against the spacecraft roll angle to remove most instrumental noise (e.g. \citealt{2022MNRAS.514...77M}). We performed a transit search on this light curve as well as 10 further-detrended light curves, each obtained by applying a different biweight time-windowed sliding filter to the first light curve. As with the TESS data, we tested 10 window sizes between 0.2 and 1.2 days. We recovered again the 6 known planets in the first 6 runs: first TOI-178\,d, then TOI-178\,g, TOI-178\,e, TOI-178\,f, TOI-178\,c, and finally TOI-178\,b. Unfortunately, the subsequent runs did not reveal any other promising transit signal.

\subsubsection{Detection limits}

To assess the detection limits of the data, we performed transit injection-and-recovery tests using the \texttt{MATRIX ToolKit}\footnote{The \texttt{MATRIX ToolKit} (Multi-phAse Transits Recovery from Injected eXoplanets ToolKit) code is available on GitHub: \url{https://github.com/PlanetHunters/tkmatrix}} (\citealt{2022zndo...6570831D}, see also \citealt{2020A&A...641A..23P,2023A&A...672A..70P}). We analysed the TESS and CHEOPS data separately, using the same input light curves as for our \texttt{SHERLOCK} transit searches for consistency. For both datasets, we explored planetary radii $R_\mathrm{p}$ between 0.5 and 3.5 $R_\oplus$ with steps of \hbox{0.2 $R_\oplus$} and orbital periods $P$ between 1 and 30 days with steps of 1 day. For each $R_\mathrm{p}-P$ combination, we injected synthetic transits at 12 random phases (i.e. 12 different values for $T_0$), thus giving a total of 5760 scenarios for each dataset. For simplicity, we assumed the impact parameters and eccentricities of the injected planets were zero. For computational cost reasons, only one detrending can be applied to the resulting light curves before trying to recover the injected transits. We chose here to use a biweight filter with a window size of 0.6 day for TESS and 0.8 day for CHEOPS. These window sizes were found to give the best results for the known planets during the \texttt{SHERLOCK} searches described above. We also masked the transits of the 6 known planets. During the transit recovery attempts, we considered a synthetic planet to be properly recovered when its epoch was found with \hbox{1 hour} accuracy and the recovered period was within 5\% of the injected period. Finally, it is worth noting that since we injected the synthetic signals into the PDCSAP light curves for TESS and the roll-angle-decorrelated light curves for CHEOPS, our results do not take into account the possible impact of these systematics corrections on the injected transits. The derived detection limits (discussed below) should therefore be considered rather optimistic (see, e.g., \citealt{2020A&A...641A..23P}, \citealt{2020MNRAS.494..750E}).

\begin{figure}[!ht]
\begin{center}
\includegraphics[width=0.49\textwidth]{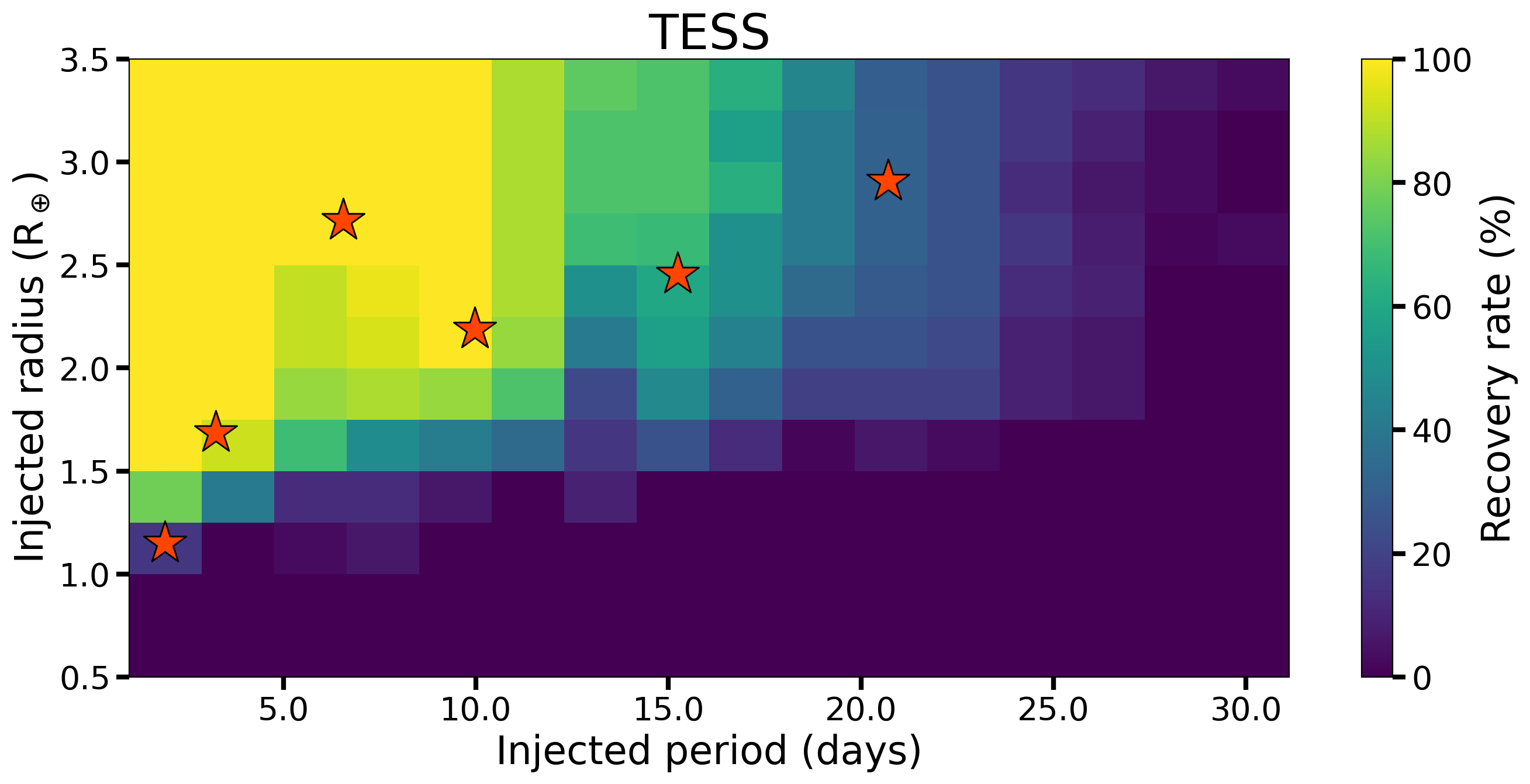}\\
\includegraphics[width=0.49\textwidth]{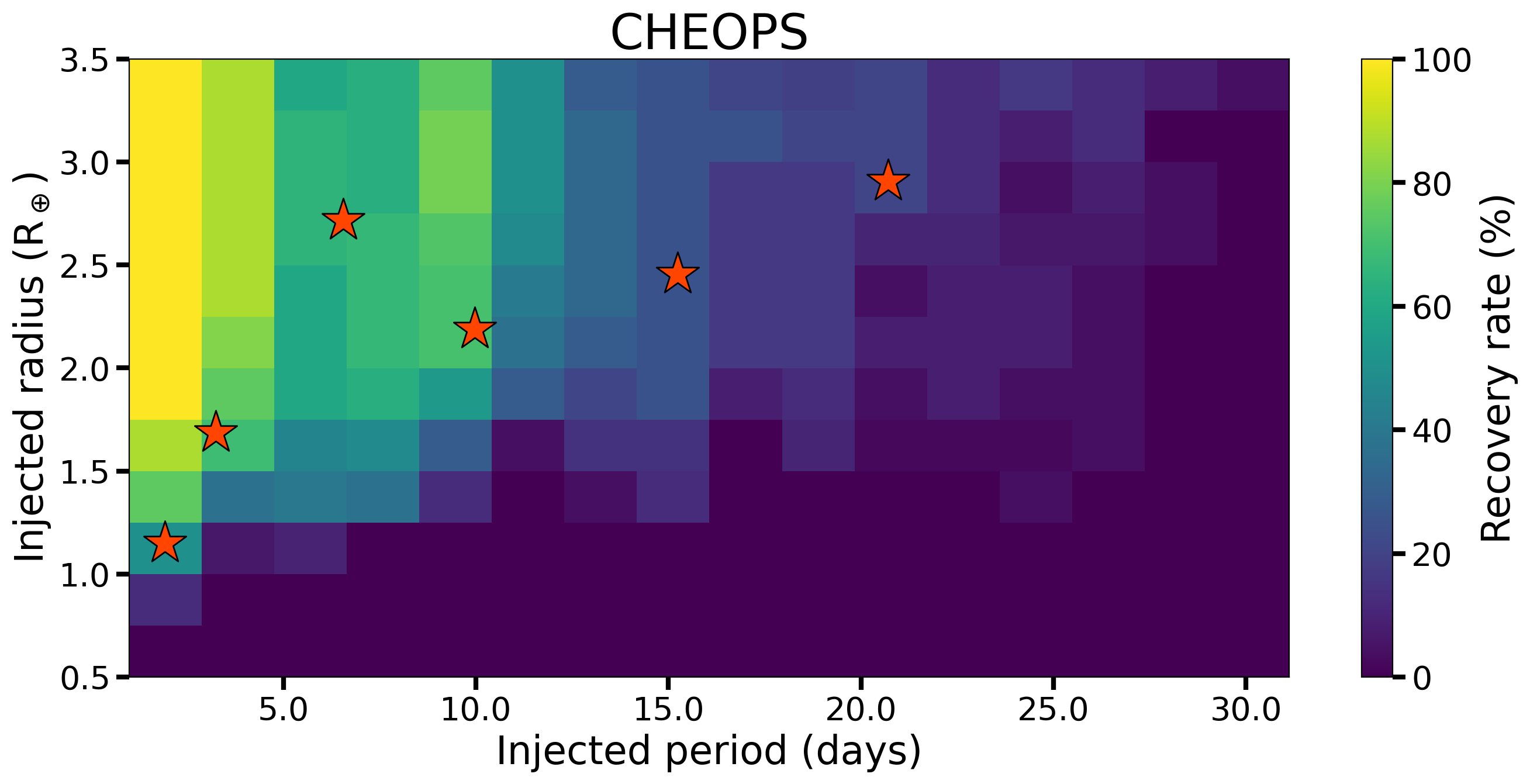}
\caption{\label{fig:TESS_detection_map} Results of the transit injection-and-recovery tests performed on the TESS (upper panel) and CHEOPS (lower panel) data to assess the detectability of potential additional transiting planets in the TOI-178 system. For both panels, the red stars show the locations of the 6 known planets in the system.} 
\end{center}
\end{figure}

Fig.~\ref{fig:TESS_detection_map} shows the two detectability maps in the $R_\mathrm{p}-P$ parameter space that we obtained for the TESS (upper panel) and CHEOPS (lower panel) data based on these transit injection-and-recovery tests. We first note that our goal here is not to compare the performances of TESS and CHEOPS. While CHEOPS's larger primary aperture size and smaller pixel scale make it a higher-precision instrument relative to TESS, the detectability of a transiting planet with a given $R_\mathrm{p}$, $P$ and $T_0$ will also strongly depend on the number of in-transit data points. In this context, it should be noted that the temporal coverage of the two datasets considered here is very different: the TESS data consist of two sets of nearly-continuous 28-day observations performed in 2018 and 2020, while the CHEOPS data include a nearly-continuous 11-day observation as well as targeted transit windows of the known planets and short observations at random times (fillers, see Sect.~\ref{sec:cheops_data}) performed in 2020 and 2021. The CHEOPS data also have a varying observing efficiency depending on the date of observation (see Sect.~\ref{sec:cheops_data}). The difference in the number of in-transit data points between the TESS and CHEOPS datasets can thus be very variable depending on the considered scenario. These considerations show that it is necessary to work on a case-by-case basis (i.e. for each $R_\mathrm{p}-P-T_{0}$ scenario) if we want to compare the performances of both instruments (see \citealt{2023AJ....165..134O} for a detailed discussion on this subject). This is not our goal here, which is rather to get a global picture of the overall detection potential of both datasets in the $R_\mathrm{p}-P$ parameter space. In this regard, Fig.~\ref{fig:TESS_detection_map} shows that additional transiting planets in the system with radii >1.75 $R_\oplus$ and orbital periods <12 days can be reasonably ruled out, as they should have been easily detected (recovery rates $\gtrsim$70\% in the TESS data and $\gtrsim$50\% in the CHEOPS data). The same planetary sizes but with orbital periods between 12 and 24 days have recovery rates ranging from $\sim$70 to 20\%. Planets of any size with orbital periods \hbox{>24 days} have recovery rates <20\%. Planets with sizes between 1.5 and \hbox{1.75 $R_\oplus$} have recovery rates $\gtrsim$50\% for orbital periods <9 days, while smaller planets with sizes between 1.0 and 1.5 $R_\oplus$ only have such reasonably good recovery rates for short orbital periods <3 days. Planets smaller than \hbox{1 $R_\oplus$} would remain undetected in the current dataset, except maybe for short orbital periods <3 days (recovery rate $\sim$12\% in the CHEOPS data, so this would be challenging).

\section{Dynamical analysis of the transit timings}
\label{sec:dynamical_analysis}

All consecutive pairs of planets in the system but the innermost one are close to a first-order MMR where $P_{\mathrm{out}}/P_{\mathrm{in}} \approx (k+1)/k$ with $k$ an integer. As none of the pairs are formally inside the 2-planet MMR, we expect TTVs over the super-period \citep{Lithwick2012}:
\begin{equation}
P_{\mathrm{c,d}} \equiv \frac{1}{|(k+1)/P_\mathrm{d} - k /P_\mathrm{c}|}\, 
\end{equation}
for planets $c$ and $d$ as an example, and similarly for the other near-resonant pairs. For TOI-178, the super-period is almost the same for all the near-resonant pairs of planets, with $P_{\mathrm{c,d}} \approx P_{\mathrm{d,e}} \approx P_{\mathrm{e,f}} \approx P_{\mathrm{f,g}} \approx 260$ days. As a result, a Laplace relation links the successive triplets of planets, leading to a slow evolution of the Laplace angles (see L21):
\begin{equation}
\begin{aligned}
\psi_1 &= 1\lambda_\mathrm{c}-4\lambda_\mathrm{d}+3\lambda_\mathrm{e}\, , \\
\psi_2 &= 2\lambda_\mathrm{d}-5\lambda_\mathrm{e}+3\lambda_\mathrm{f}\, , \\
\psi_3 &= 1\lambda_\mathrm{e}-3\lambda_\mathrm{f}+2\lambda_\mathrm{g}
\end{aligned}
\end{equation}
where $\lambda_\mathrm{i}$ is the mean longitude of planet $i$. The evolution of these angles can also produce TTVs on much longer timescales, although according to the mass estimates obtained using radial velocities (see L21), these effects should only start to show with at least 4 years of baseline. Finally, the relative proximity of the planets can also generate a high-frequency chopping signal \citep{DeAg2015}.

\begin{figure}[!ht]
\begin{center}
\includegraphics[width=0.49\textwidth]{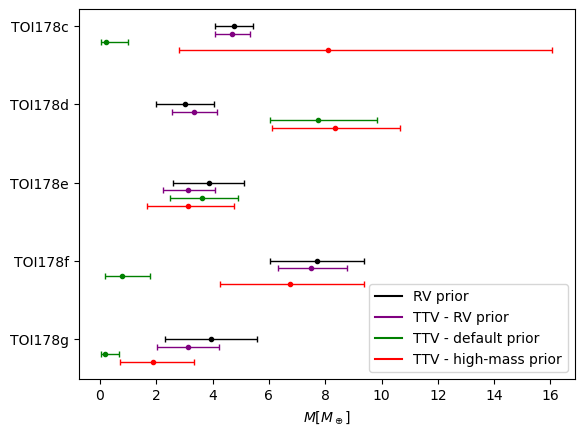}\\
\includegraphics[width=0.49\textwidth]{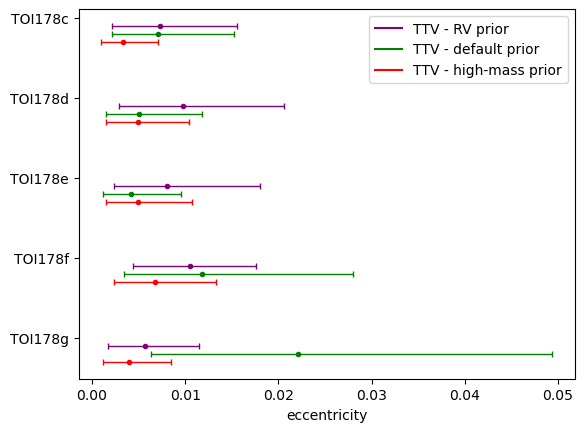}\\
\caption{\label{fig:masscomp} Mass (upper panel) and eccentricity (lower panel) posteriors derived from our dynamical analysis of the TTVs. The corresponding values are given in Table \ref{tab:masscomp}. In the upper panel, the radial velocity prior based on the RV measurements presented in L21 is shown in black, with $1\sigma$ error bars. The coloured error bars show the median and .16-.84 quantiles confidence level of the mass posteriors obtained from our TTV fit using three different sets of priors.}
\end{center}
\end{figure}

\begin{figure*}
    \centering
    \includegraphics[width=0.49\textwidth]{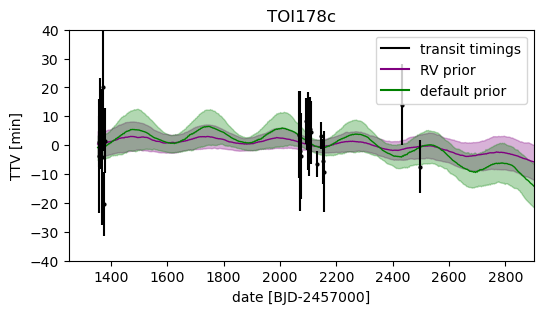}
    \includegraphics[width=0.49\textwidth]{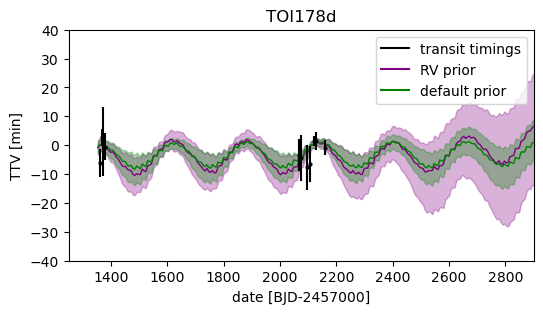}
    \includegraphics[width=0.49\textwidth]{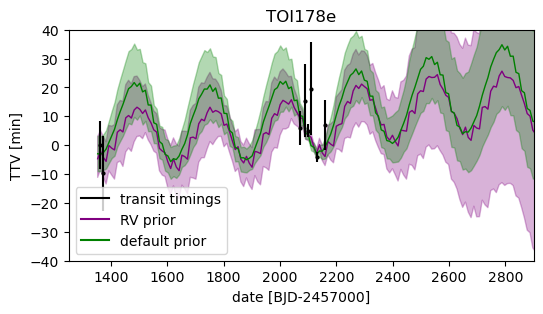}
    \includegraphics[width=0.49\textwidth]{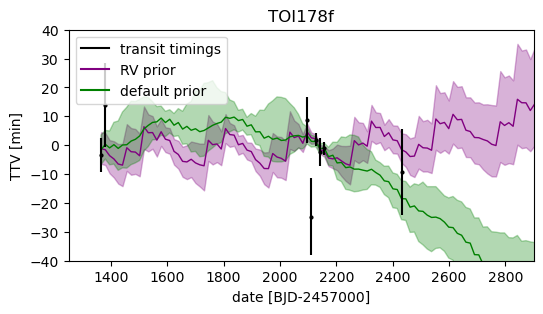}
    \includegraphics[width=0.49\textwidth]{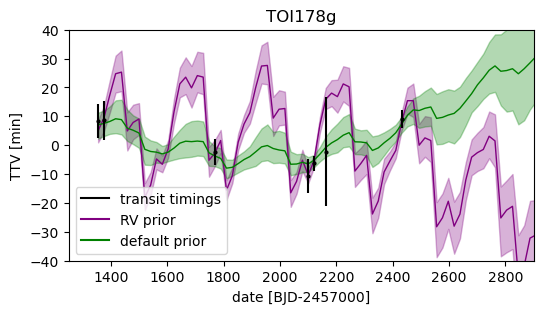}
    \caption{For each planet, the black points with error bars show the measured TTVs in minutes. For each planet, we also show the median and 1-sigma envelope for the TTVs propagated from 300 randomly-selected samples of the posterior obtained with the \textit{RV} (purple) and \textit{default} (green) mass priors.}
    \label{fig:ttvs}
\end{figure*}

We study here the transit timings reported in Table \ref{tab:timings} for the five outer planets. L21 predicted TTV peak-to-peak amplitudes ranging from several minutes for the inner planets to a few tens of minutes for the outer ones. Since the timing uncertainties of the inner planets are comparable to the expected signal, we proceed in two steps. First, we check whether the observed transit timings are consistent with the RV masses and if some constraints on the eccentricities can be obtained by combining the two. As a second step, we then assess the constraints on the masses that can be derived from the TTVs alone. 

We fit the TTVs using the code presented in \cite{Leleu2021b}: the transit timings are estimated using the {\ttfamily TTVfast} algorithm \citep{DeAgHo2014} and the \texttt{samsam}\footnote{\url{https://gitlab.unige.ch/Jean-Baptiste.Delisle/samsam}} MCMC algorithm \citep[see][]{Delisle2018} is used to sample the posteriors. As L21 showed variations in the projected orbital inclination of only about 0.1 degree between the outer planets (Fig. 8 of L21), we assume in this study that the system is coplanar. The mean longitudes, periods, arguments of periastron, and eccentricities of the planets have uniform priors. For our first test, the mass priors are Gaussian with the respective mean and standard deviation based on the RV posteriors presented in L21. We call this setup the \textit{RV} prior. The mass and eccentricity posteriors of this fit are shown in Fig. \ref{fig:masscomp} and given in Table \ref{tab:masscomp} (together with the \textit{RV} mass priors for comparison). Fig. \ref{fig:masscomp} shows that, for each planet, the mass posterior of the TTV fit is 1-$\sigma$ consistent with the \textit{RV} prior, implying that the observed TTVs are indeed compatible with the RV masses. When using the \textit{RV} mass priors, the TTVs allow us to constrain the eccentricities. L21 set the eccentricities to 0 as the available RVs did not allow a precise measurement, and the stability analysis of the system showed that the eccentricities had to be of a few percent at most. Here, the posteriors of the TTV fit obtained with the \textit{RV} mass priors explore eccentricities that are consistent with both the masses determined from the RVs (all mass posteriors are 1-$\sigma$ consistent with their priors) and the TTV signals. We report the derived eccentricities in Tables \ref{tab:transit_parameters1} (planets $b$, $c$, and $d$) and \ref{tab:transit_parameters2} (planets $e$, $f$, and $g$). The 0.84 quantiles of the eccentricities are below 0.021 for all planets, which is consistent with the stability study of the system presented in L21.

As a second step, we check which constraints on the masses can be obtained from the TTVs alone. The main TTV signal whose period is the aforementioned super-period is degenerate between the planetary masses and eccentricities \citep{Lithwick2012}. The observation of other TTV harmonics, such as the chopping signal or the resonant evolution of the Laplace angle is thus necessary to constrain the planetary masses. Following \cite{HaLi2016,HaLi2017}, we hence fit the data with different mass priors to test the robustness of the retrieved masses. The \textit{default} prior is log-uniform in planetary masses, while the \textit{high-mass} prior is uniform in planetary masses. Posteriors that we obtained using these priors are also shown in Fig. \ref{fig:masscomp}. We quantify the robustness of the mass determination using the parameter: 
\begin{equation}
\Delta_M=\left| \frac{M_{\text{default}}^{.5}-M_{\text{high-mass}}^{.5}}{M_{\text{default}}^{.84}-M_{\text{default}}^{.5}}   \right|
\end{equation}
where $M_{\text{default}}^{.5}$ is the quantile at .5 of the \textit{default} mass posterior, and similarly for other quantities. The robustness mass criterion from \cite{HaLi2017} requires that $\Delta_M < 1$ (note that their criterion is even more conservative as their high-mass prior also contains a log-uniform eccentricity prior, while in our case the eccentricity prior is uniform for all posteriors). The values that we obtained for $\Delta_M$, reported in Table \ref{tab:masscomp}, imply that the mass estimations of planets $c$, $f$, and $g$ are highly degenerate ($\Delta_M>3$ for the three of them). The test finds robust masses for planets $d$ and $e$. For TOI-178\,e, the medians of both the \textit{default} and \textit{high-mass} posteriors are within 1-$\sigma$ of the \textit{RV} prior, while for TOI-178\,d, the medians of the two posteriors are both outside the 1-$\sigma$ interval of the \textit{RV} prior.

However, the mean log-Likelihood computed for each posterior differs by less than 1 across all three. It implies that the three solutions explain the data equally well and that the various priors explore different parts of the underlying degeneracies. As a result, we recommend keeping for now the RV mass estimates presented in L21 as nominal values for the masses of the planets, as we deem the current TTV mass posteriors too much prior-dependent. Nonetheless, the apparent mass shift between the \textit{RV} prior and the higher mass posterior found with both the \textit{default} and \textit{high-mass} priors for TOI-178\,d highlights the importance of continuing to monitor the system in the future, both in RVs and TTVs. Indeed, TOI-178\,d had been found by L21 to have a surprisingly low density based on its RV mass estimate (Sect. \ref{sec:intro}). Differences in planet density between RV- and TTV-characterised systems have been discussed in numerous studies over the last decade \citep[e.g.][]{HaLi2017,2017ApJ...839L...8M,Leleu2023}. TOI-178 offers the rare opportunity to compare the two techniques for the same system. 

Fig. \ref{fig:ttvs} shows the measured transit timings, as well as the TTV posteriors obtained with the \textit{RV} (purple) and \textit{default} (green) mass priors. We note that all current TTV measurements are well explained by the known planets of the system. The TTV signal over the $\sim$260d super-period is clearly visible in the posteriors of all planets. For the \textit{RV} prior, a strong chopping signal is also visible for the two outer planets, while for the \textit{default} prior, the TTVs are explained mainly through the slow evolution of the Laplace angles. A two-year projection after the last observed transit shows that future observations should be able to distinguish between these solutions. 

Finally, we note that a photodynamical analysis of the light curves, where the gravitational interactions between the planets are taken into account at the stage of the light curve modelling \citep[e.g.][]{RaHo2010, 2018MNRAS.478..460A, 2022A&A...663A.134A}, could help to better constrain the orbital parameters and masses. In particular, it has been shown that a photodynamical analysis can help to reduce the mass-eccentricity degeneracy when compared to the fit of pre-extracted transit timings, in particular for planets in the super-Earth to mini-Neptune range \citep{Leleu2023}. Such a photodynamical analysis will be performed in an upcoming paper (Leleu et al. 2023, in prep.).

\section{Conclusions}
\label{sec:conclusion}

In this work, we presented a detailed photometric study of the TOI-178 system, based on 40 new CHEOPS visits, one new TESS sector, as well as previously published data. We first performed a global analysis of the 100 transits contained in these data. This enabled us to significantly refine the transit parameters of the six TOI-178 planets, most notably their radii, for which we obtain relative precisions $\lesssim$3\%, with the exception of the smallest planet $b$ for which the precision is 5.1\%. We also used our extensive photometric dataset to place constraints on the radii and orbital periods of potential additional transiting planets in the system.

As part of our study, we also performed a first dynamical analysis of the TTVs measured for the five outer planets ($c$ to $g$), testing different priors for their masses to assess the robustness of the derived solution. We found that the mass posteriors are very prior-dependent. On one hand, when fitting the TTVs with mass priors based on the previously-published RVs, we find masses that are consistent with the RVs, and eccentricities that are all below 0.02, as expected from stability requirements.
On the other hand, when fitting the TTVs with uniform or log-uniform (RV-independent) mass priors, we find mass estimates that are highly degenerate for planets $c$, $f$, and $g$; consistent with the RVs for planet $e$; and higher than the RV mass for planet $d$. We note that this latter planet had been found by L21 to have a surprisingly low density based on its RV mass estimate. Since the masses derived from the current TTV dataset are very prior-dependent, we recommend keeping for now the RV mass estimates presented in L21 as nominal values for the masses of the planets. Altogether, this first TTV study highlights the importance of continuing to monitor the system in the future, both in RVs and TTVs. In this context, further TTV measurements with CHEOPS and TESS (TOI-178 will be observed again in Sector 69), covering a longer temporal baseline, should help to break the degeneracies and improve our understanding of this benchmark planetary system.
 
\begin{acknowledgements}
CHEOPS is an ESA mission in partnership with Switzerland with important contributions to the payload and the ground segment from Austria, Belgium, France, Germany, Hungary, Italy, Portugal, Spain, Sweden, and the United Kingdom. The CHEOPS Consortium would like to gratefully acknowledge the support received by all the agencies, offices, universities, and industries involved. Their flexibility and willingness to explore new approaches were essential to the success of this mission. 
The Belgian participation to CHEOPS has been supported by the Belgian Federal Science Policy Office (BELSPO) in the framework of the PRODEX Program, and by the University of Liège through an ARC grant for Concerted Research Actions financed by the Wallonia-Brussels Federation. 
L.D. is an F.R.S.-FNRS Postdoctoral Researcher. 
This work has been carried out within the framework of the NCCR PlanetS supported by the Swiss National Science Foundation under grants 51NF40\_182901 and 51NF40\_205606. 
A.Br. was supported by the SNSA. 
M.G. is an F.R.S.-FNRS Senior Research Associate. 
ACC acknowledges support from STFC consolidated grant numbers ST/R000824/1 and ST/V000861/1, and UKSA grant number ST/R003203/1.
This work was also partially supported by a grant from the Simons Foundation (PI Queloz, grant number 327127).
V.V.G. is an F.R.S-FNRS Research Associate. 
A.C.C. and T.G.W. acknowledge support from STFC consolidated grant numbers ST/R000824/1 and ST/V000861/1, and UKSA grant number ST/R003203/1. 
Y.A. and M.J.H. acknowledge the support of the Swiss National Fund under grant 200020\_172746. 
We acknowledge support from the Spanish Ministry of Science and Innovation and the European Regional Development Fund through grants ESP2016-80435-C2-1-R, ESP2016-80435-C2-2-R, PGC2018-098153-B-C33, PGC2018-098153-B-C31, ESP2017-87676-C5-1-R, MDM-2017-0737 Unidad de Excelencia Maria de Maeztu-Centro de Astrobiología (INTA-CSIC), as well as the support of the Generalitat de Catalunya/CERCA programme. The MOC activities have been supported by the ESA contract No. 4000124370. 
S.C.C.B. acknowledges support from FCT through FCT contracts nr. IF/01312/2014/CP1215/CT0004. 
X.B., S.C., D.G., M.F. and J.L. acknowledge their role as ESA-appointed CHEOPS science team members. 
L.Bo., G.Br., V.Na., I.Pa., G.Pi., R.Ra., G.Sc., V.Si., and T.Zi. acknowledge support from CHEOPS ASI-INAF agreement n. 2019-29-HH.0.
P.E.C. is funded by the Austrian Science Fund (FWF) Erwin Schroedinger Fellowship, program J4595-N.
This project was supported by the CNES. 
This work was supported by FCT - Fundação para a Ciência e a Tecnologia through national funds and by FEDER through COMPETE2020 - Programa Operacional Competitividade e Internacionalizacão by these grants: UID/FIS/04434/2019, UIDB/04434/2020, UIDP/04434/2020, PTDC/FIS-AST/32113/2017 \& POCI-01-0145-FEDER- 032113, PTDC/FIS-AST/28953/2017 \& POCI-01-0145-FEDER-028953, PTDC/FIS-AST/28987/2017 \& POCI-01-0145-FEDER-028987. O.D.S.D. is supported in the form of work contract (DL 57/2016/CP1364/CT0004) funded by national funds through FCT. 
B.-O.D. acknowledges support from the Swiss State Secretariat for Education, Research and Innovation (SERI) under contract number MB22.00046. 
This project has received funding from the European Research Council (ERC) under the European Union's Horizon 2020 research and innovation programme (project {\sc Four Aces}, grant agreement No 724427). It has also been carried out in the frame of the National Centre for Competence in Research PlanetS supported by the Swiss National Science Foundation (SNSF). D.E. acknowledges financial support from the Swiss National Science Foundation for project 200021\_200726. 
M.F. gratefully acknowledges the support of the Swedish National Space Agency (DNR 65/19, 174/18). 
D.G. gratefully acknowledges financial support from the CRT foundation under Grant No. 2018.2323 `Gaseous or rocky? Unveiling the nature of small worlds'. 
S.H. gratefully acknowledges CNES funding through the grant 837319. 
K.G.I. is the ESA CHEOPS Project Scientist and is responsible for the ESA CHEOPS Guest Observers Programme. She does not participate in, or contribute to, the definition of the Guaranteed Time Programme of the CHEOPS mission through which observations described in this paper have been taken, nor to any aspect of target selection for the programme. 
This work was granted access to the HPC resources of MesoPSL financed by the Region Ile de France and the project Equip@Meso (reference ANR-10-EQPX-29-01) of the programme Investissements d'Avenir supervised by the Agence Nationale pour la Recherche. 
M.L. acknowledges support of the Swiss National Science Foundation under grant number PCEFP2\_194576.
R.L. acknowledges funding from University of La Laguna through the Margarita Salas Fellowship from the Spanish Ministry of Universities ref. UNI/551/2021-May 26, and under the EU Next Generation funds. 
P.M. acknowledges support from STFC research grant number ST/M001040/1. 
I.Ri. acknowledges support from the Spanish Ministry of Science and Innovation and the European Regional Development Fund through grant PGC2018-098153-B- C33, as well as the support of the Generalitat de Catalunya/CERCA programme. 
S.G.S. acknowledge support from FCT through FCT contract nr. CEECIND/00826/2018 and POPH/FSE (EC). 
Gy.M.Sz. acknowledges the support of the Hungarian National Research, Development and Innovation Office (NKFIH) grant K-125015, a PRODEX Experiment Agreement No. 4000137122, the Lend\"ulet LP2018-7/2021 grant of the Hungarian Academy of Science and the support of the city of Szombathely. 
N.A.W. acknowledges UKSA grant ST/R004838/1. Funding for the TESS mission is provided by the NASA's Science Mission Directorate. We acknowledge the use of public TESS data from
pipelines at the TESS Science Office and at the TESS Science Processing Operations
Center. Resources supporting this work were provided by the NASA High-End Computing (HEC) Program through the NASA Advanced Supercomputing (NAS) Division at Ames Research Center for the production of the SPOC data products. This paper includes data collected by the TESS mission that are publicly available from the Mikulski Archive for Space Telescopes (MAST). 
We thank the anonymous referee for taking the time to review our work and for her/his valuable suggestions.

\end{acknowledgements}

\bibliographystyle{aa}
\bibliography{biblio}

\onecolumn
\begin{appendix}

\section{Data: CHEOPS visits}

\begin{table*}[hbt!]
\begin{small}
    \centering
    \caption{Log of the CHEOPS observations. The file key (first column) is a unique identifier that can be used to retrieve the data from the CHEOPS archive. The efficiency (sixth column) represents the ratio between the amount of science observing time in a visit (excluding the interruptions due to Earth occultations or SAA crossings) and the total duration of the visit (including interruptions). The last two columns give the median absolute deviation (MAD) of the difference between two consecutive data points of the light curve for the CHEOPS Data Reduction Pipeline (DRP) and the PSF photometry pipeline (PIPE).} 
    \begin{tabular}{cccccccc}
    \hline\hline
    File key & UTC start & UTC end & Content & $N_{\mathrm{frames}}$ & Efficiency & DRP MAD & PIPE MAD \\
    \vspace{0.1cm}
     & & & & & (\%) & (ppm) & (ppm) \\
    \hline
    \texttt{PR100031\_TG028901} & 2020-07-21 11:01 & 2020-07-21 14:28 & Filler & 112 & 54.1 & 632 & 485 \\
    \texttt{PR100031\_TG029001} & 2020-07-25 01:59 & 2020-07-25 05:13 & Filler & 109 & 56.2 & 598 & 550 \\
    \texttt{PR100031\_TG030201}$^{(a)}$ & 2020-08-04 22:10 & 2020-08-09 01:57 & b ($\times$2), c, d, e & 3253 & 54.3 & 662 & 590 \\
    \texttt{PR100031\_TG030301}$^{(a)}$ & 2020-08-09 02:47 & 2020-08-15 22:50 & b ($\times$4), c ($\times$2), d & 5590 & 56.8 & 665 & 590 \\
    \texttt{PR100031\_TG030701}$^{(a)}$ & 2020-09-07 08:06 & 2020-09-07 21:27 & e, g & 559 & 69.7 & 740 & 638 \\
    \texttt{PR100031\_TG032101} & 2020-09-20 21:02 & 2020-09-21 01:54 & Filler & 257 & 87.7 & 741 & 596 \\
    \texttt{PR100031\_TG031001} & 2020-09-22 21:02 & 2020-09-23 00:15 & Filler & 165 & 85.0 & 648 & 583 \\
    \texttt{PR100031\_TG031801} & 2020-09-23 00:26 & 2020-09-23 05:18 & Filler & 283 & 96.5 & 747 & 532 \\
    \texttt{PR100031\_TG032201} & 2020-09-28 03:51 & 2020-09-28 12:17 & b, d, g & 441 & 86.9 & 759 & 657 \\
    \texttt{PR100031\_TG031101} & 2020-09-29 06:23 & 2020-09-29 11:24 & Filler & 253 & 83.7 & 631 & 638 \\
    \texttt{PR100031\_TG031201} & 2020-10-01 07:32 & 2020-10-01 10:45 & Filler & 160 & 82.4 & 1016 & 650 \\
    \texttt{PR100031\_TG031401} & 2020-10-03 10:24 & 2020-10-03 13:37 & Filler & 180 & 92.7 & 879 & 664 \\
    \texttt{PR100031\_TG033301}$^{(a)}$ & 2020-10-03 18:51 & 2020-10-04 02:51 & b, f & 424 & 88.1 & 750 & 593 \\
    \texttt{PR100031\_TG031301} & 2020-10-04 17:02 & 2020-10-04 19:59 & Filler & 161 & 90.4 & 798 & 583 \\
    \texttt{PR100031\_TG033001} & 2020-10-04 20:26 & 2020-10-05 05:09 & d & 480 & 91.6 & 686 & 550 \\    
    \texttt{PR100031\_TG033302} & 2020-10-05 16:42 & 2020-10-05 23:56 & b & 379 & 87.1 & 825 & 574 \\  
    \texttt{PR100031\_TG033101} & 2020-10-07 06:05 & 2020-10-07 15:26 & c, e & 497 & 88.4 & 762 & 604 \\  
    \texttt{PR100031\_TG033303} & 2020-10-07 15:37 & 2020-10-07 22:51 & b & 372 & 85.5 & 844 & 613 \\    
    \texttt{PR100031\_TG031901} & 2020-10-08 22:04 & 2020-10-09 02:56 & Filler & 273 & 93.1 & 687 & 602 \\
    \texttt{PR100031\_TG032001} & 2020-10-09 08:10 & 2020-10-09 13:02 & Filler & 254 & 86.6 & 737 & 571 \\
    \texttt{PR100031\_TG033304} & 2020-10-09 13:13 & 2020-10-09 20:20 & b & 399 & 93.2 & 755 & 593 \\    
    \texttt{PR100031\_TG033901} & 2020-10-12 18:09 & 2020-10-12 23:27 & Filler & 236 & 73.9 & 782 & 609 \\   
    \texttt{PR100031\_TG035101} & 2020-10-19 02:36 & 2020-10-19 10:20 & b, f & 333 & 71.6 & 734 & 617 \\      
    \texttt{PR100031\_TG033902} & 2020-10-20 00:21 & 2020-10-20 05:13 & Filler & 218 & 74.4 & 927 & 633 \\  
    \texttt{PR100031\_TG034001} & 2020-10-21 11:19 & 2020-10-21 14:32 & Filler & 131 & 67.5 & 657 & 464 \\    
    \texttt{PR100031\_TG033903} & 2020-10-22 11:52 & 2020-10-22 16:41 & Filler & 193 & 66.5 & 796 & 605 \\  
    \texttt{PR100031\_TG034201} & 2020-10-23 15:18 & 2020-10-23 22:52 & c & 262 & 57.6 & 759 & 669 \\ 
    \texttt{PR100031\_TG034002} & 2020-10-24 20:59 & 2020-10-24 23:52 & Filler & 97 & 55.7 & 892 & 593 \\     
    \texttt{PR100031\_TG033904} & 2020-10-26 06:00 & 2020-10-26 10:53 & Filler & 174 & 59.4 & 595 & 469 \\    
    \texttt{PR100031\_TG034003} & 2020-10-28 01:30 & 2020-10-28 04:43 & Filler & 115 & 59.3 & 756 & 593 \\
    \texttt{PR100031\_TG035601} & 2020-10-28 17:17 & 2020-10-29 00:16 & b & 115 & 59.3 & 841 & 736 \\  
    \texttt{PR100031\_TG033905} & 2020-10-29 20:30 & 2020-10-30 01:33 & Filler & 153 & 50.3 & 712 & 581 \\
    \texttt{PR100031\_TG035701} & 2020-10-30 02:16 & 2020-10-30 09:46 & c & 285 & 63.2 & 723 & 647 \\    
    \texttt{PR100031\_TG035702} & 2020-11-02 08:03 & 2020-11-02 16:07 & c & 274 & 56.5 & 715 & 634 \\  
    \texttt{PR100031\_TG036101} & 2020-11-03 07:32 & 2020-11-03 14:19 & b, f & 234 & 57.3 & 627 & 614 \\ 
    \texttt{PR100031\_TG036401} & 2020-11-06 04:41 & 2020-11-06 14:02 & e & 289 & 51.4 & 644 & 517 \\  
    \texttt{PR100031\_TG036301} & 2020-11-06 14:13 & 2020-11-06 22:56 & d & 240 & 45.8 & 675 & 558 \\   
    \texttt{PR100031\_TG035602} & 2020-11-07 07:09 & 2020-11-07 13:55 & b & 234 & 57.5 & 714 & 626 \\           
    \texttt{PR100031\_TG036201} & 2020-11-08 14:11 & 2020-11-08 22:13 & g & 218 & 45.1 & 766 & 577 \\    
    \texttt{PR100031\_TG043601} & 2021-08-04 09:14 & 2021-08-04 18:48 & b, c, f & 306 & 53.2 & 691 & 618 \\        
    \texttt{PR100031\_TG043701} & 2021-08-04 19:00 & 2021-08-05 05:53 & c, g & 360 & 55.0 & 740 & 668 \\       
    \texttt{PR100031\_TG043201} & 2021-10-08 06:52 & 2021-10-08 14:01 & b, c & 378 & 87.9 & 682 & 602 \\      
    \texttt{PR100031\_TG043202} & 2021-10-15 22:47 & 2021-10-16 05:56 & b & 343 & 79.7 & 650 & 552 \\  
    \texttt{PR100031\_TG043203} & 2021-10-17 21:29 & 2021-10-18 04:38 & b & 304 & 70.7 & 679 & 587 \\      
    \hline\hline
    \end{tabular}
    \textbf{Notes.} $^{(a)}$ Data previously presented in L21.
    \label{tab:CHEOPS_visits}
\end{small}
\end{table*}

\clearpage

\section{Global transit analysis: supplementary material}

\subsection{Photometric baseline models and error scaling factors}

\begin{longtable}{c c c c c c c c c c}
\hline
\hline
Date & Facility & Planet(s) & Epoch(s) & $N_{\rm points}$ & $T_{\rm exp}$ & Baseline model & Residual RMS & $\beta_w$ & $\beta_r$ \\
(UT) & & & & & (s) & & (\%) & & \\
\hline
\endfirsthead
\multicolumn{9}{c}{\tablename\ \thetable{} -- \textit{continued from previous page}} \\
\hline
Date & Facility & Planet(s) & Epoch(s) & $N_{\rm points}$ & $T_{\rm exp}$ & Baseline model & Residual RMS & $\beta_w$ & $\beta_r$ \\
(UT) & & & & & (s) & & (\%) & & \\
\hline
\endhead
\hline \multicolumn{9}{c}{\textit{Continued on next page}} \\ 
\endfoot
\hline \hline
\caption{Photometric baseline models, residual RMS, and error scaling factors $\beta_w$ and $\beta_r$ for each transit light curve used in our global transit analysis (see Sect. \ref{sec:analysis}). For the baseline function, $o$ is a simple constant (to account for any out-of-transit flux offset), sp($r$) is a cubic spline against the spacecraft roll angle $r$, and p($\alpha$) denotes a first-order polynomial function of the parameter $\alpha$, with $\alpha$ that can be $t=$ 
time, $b=$ background, $a=$ airmass, $x$ and $y=$ $x$- and $y$- position of the target on the detector, or $ttt=$ telescope tube temperature. Epochs are relative to the updated mean transit ephemerides given in Tables \ref{tab:transit_parameters1} (planets $b$, $c$, and $d$) and \ref{tab:transit_parameters2} (planets $e$, $f$, and $g$).}
\label{tab:baselines}\\
\endlastfoot
2018-09-23 & TESS & g & -26 & 265 & 120 & $o$ & 0.118 & 0.94 & 1.74 \\
2018-08-24 & TESS & b & -301 & 196 & 120 & $o$ & 0.118 & 0.94 & 1.29 \\
2018-08-26 & TESS & c & -176 & 218 & 120 & $o$ & 0.121 & 0.96 & 1.23 \\
2018-08-26 & TESS & b & -300 & 195 & 120 & p($t$) & 0.116 & 0.92 & 1.14 \\
2018-08-28 & TESS & b & -299 & 187 & 120 & $o$ & 0.123 & 0.97 & 1.48 \\
2018-08-28 & TESS & c & -175 & 228 & 120 & $o$ & 0.126 & 1.00 & 1.08 \\
2018-08-29 & TESS & d & -61 & 247 & 120 & p($t$) & 0.127 & 1.01 & 1.08 \\
2018-08-30 & TESS & b & -298 & 197 & 120 & $o$ & 0.108 & 0.86 & 1.00 \\
2018-08-31 & TESS & b, c & -297, -174 & 229 & 120 & $o$ & 0.119 & 0.94 & 1.15 \\
2018-09-01 & TESS & e & -40 & 293 & 120 & $o$ & 0.114 & 0.90 & 1.09 \\
2018-09-02 & TESS & b & -296 & 195 & 120 & $o$ & 0.118 & 0.93 & 1.00 \\
2018-09-03 & TESS & f & -35 & 283 & 120 & $o$ & 0.125 & 0.97 & 1.13 \\
2018-09-04 & TESS & c & -173 & 227 & 120 & $o$ & 0.118 & 0.92 & 1.35 \\
2018-09-04 & TESS & b & -295 & 193 & 120 & $o$ & 0.135 & 1.05 & 1.28 \\
2018-09-05 & TESS & d & -60 & 235 & 120 & $o$ & 0.108 & 0.84 & 1.07 \\
2018-09-07 & TESS & c & -172 & 229 & 120 & $o$ & 0.133 & 1.05 & 1.65 \\
2018-09-08 & TESS & b & -293 & 202 & 120 & $o$ & 0.118 & 0.94 & 1.33 \\
2018-09-10 & TESS & b & -292 & 196 & 120 & $o$ & 0.129 & 1.02 & 1.36 \\
2018-09-10 & TESS & c & -171 & 225 & 120 & $o$ & 0.119 & 0.94 & 1.33 \\
2018-09-11 & TESS & e & -39 & 296 & 120 & $o$ & 0.126 & 1.00 & 1.79 \\
2018-09-11 & TESS & d & -59 & 239 & 120 & $o$ & 0.120 & 0.95 & 1.47 \\
2018-09-12 & TESS & b & -291 & 190 & 120 & $o$ & 0.111 & 0.88 & 1.71 \\
2018-09-13 & TESS & g & -25 & 265 & 120 & $o$ & 0.122 & 0.97 & 1.65 \\
2018-09-13 & TESS & c & -170 & 226 & 120 & $o$ & 0.122 & 0.97 & 1.00 \\
2018-09-14 & TESS & b & -290 & 186 & 120 & $o$ & 0.111 & 0.88 & 1.47 \\
2018-09-16 & TESS & b & -289 & 186 & 120 & $o$ & 0.132 & 1.04 & 1.59 \\
2018-09-17 & TESS & c & -169 & 206 & 120 & $o$ & 0.117 & 0.92 & 1.65 \\
2018-09-18 & TESS & b & -288 & 195 & 120 & $o$ & 0.121 & 0.94 & 1.20 \\
2018-09-18 & TESS & d & -58 & 173 & 120 & $o$ & 0.129 & 1.00 & 1.02 \\
2018-09-18 & TESS & f & -34 & 268 & 120 & $o$ & 0.130 & 1.00 & 1.11 \\
2019-09-11 & NGTS1 & b & -101 & 2321 & 10 & p($a$) & 0.639 & 0.72 & 1.13 \\
2019-09-11 & NGTS2 & b & -101 & 2332 & 10 & p($a$) & 0.627 & 0.75 & 1.23 \\
2019-09-11 & NGTS3 & b & -101 & 2326 & 10 & p($a$) & 0.634 & 0.76 & 1.48 \\
2019-09-11 & NGTS4 & b & -101 & 2325 & 10 & p($a$) & 0.633 & 0.74 & 1.01 \\
2019-09-11 & NGTS5 & b & -101 & 2330 & 10 & p($a$) & 0.637 & 0.75 & 1.46 \\
2019-09-11 & NGTS6 & b & -101 & 2309 & 10 & p($a$) & 0.639 & 0.78 & 1.96 \\
2019-10-12 & NGTS1 & g & -6 & 1835 & 10 & p($a$) & 0.578 & 0.60 & 2.74 \\
2019-10-12 & NGTS2 & g & -6 & 1828 & 10 & p($a$) & 0.581 & 0.60 & 1.46 \\
2019-10-12 & NGTS3 & g & -6 & 1838 & 10 & p($a$) & 0.616 & 0.61 & 1.11 \\
2019-10-12 & NGTS4 & g & -6 & 1823 & 10 & p($a$) & 0.553 & 0.56 & 1.42 \\
2019-10-12 & NGTS5 & g & -6 & 1836 & 10 & p($a$) & 0.574 & 0.60 & 1.00 \\
2019-10-12 & NGTS6 & g & -6 & 1849 & 10 & p($a$) & 0.567 & 0.59 & 2.23 \\
2019-10-12 & NGTS7 & g & -6 & 1831 & 10 & p($a$) & 0.578 & 0.59 & 1.65 \\
2020-08-05 & CHEOPS & b & 71 & 206 & 60 & sp($r$) + $o$ & 0.057 & 1.01 & 1.28 \\
2020-08-06 & CHEOPS & c, d & 44, 47 & 404 & 60 & sp($r$) + $o$ & 0.057 & 1.00 & 2.49 \\
2020-08-07 & CHEOPS & b & 72 & 190 & 60 & sp($r$) + $o$ & 0.057 & 1.00 & 1.55 \\
2020-08-08 & CHEOPS & e & 31 & 331 & 60 & sp($r$) + $o$ & 0.057 & 1.00 & 2.11 \\
2020-08-09 & CHEOPS & b & 73 & 196 & 60 & sp($r$) + $o$ & 0.060 & 1.07 & 1.83 \\
2020-08-10 & CHEOPS & c & 45 & 185 & 60 & sp($r$) + $o$ & 0.056 & 1.00 & 1.07 \\
2020-08-11 & CHEOPS & b & 74 & 187 & 60 & sp($r$) + $o$ & 0.056 & 1.00 & 1.92 \\
2020-08-12 & CHEOPS & b & 75 & 275 & 60 & sp($r$) + $o$ & 0.056 & 1.01 & 2.23 \\
2020-08-13 & CHEOPS & c, d & 46, 48 & 374 & 60 & sp($r$) + p($b$) & 0.058 & 1.04 & 2.14 \\
2020-08-15 & CHEOPS & b & 76 & 279 & 60 & sp($r$) + $o$ & 0.056 & 1.00 & 1.13 \\
2020-08-28 & TESS & b, e & 83, 33 & 368 & 120 & $o$ & 0.124 & 0.99 & 1.19 \\
2020-08-29 & TESS & c & 51 & 241 & 120 & $o$ & 0.124 & 0.99 & 1.00 \\
2020-08-30 & TESS & b & 84 & 207 & 120 & $o$ & 0.119 & 0.94 & 1.09 \\
2020-09-01 & TESS & b & 85 & 208 & 120 & $o$ & 0.122 & 0.97 & 1.26 \\
2020-09-01 & TESS & c & 52 & 240 & 120 & $o$ & 0.115 & 0.91 & 1.00 \\
2020-09-02 & TESS & d & 51 & 236 & 120 & $o$ & 0.131 & 1.04 & 1.08 \\
2020-09-03 & TESS & b & 86 & 208 & 120 & $o$ & 0.110 & 0.87 & 1.07 \\
2020-09-03 & TESS & f & 13 & 288 & 120 & $o$ & 0.112 & 0.88 & 1.32 \\
2020-09-04 & TESS & b, c & 87, 53 & 361 & 120 & $o$ & 0.120 & 0.93 & 1.50 \\
2020-09-07 & CHEOPS & e, g & 34, 10 & 555 & 60 & sp($r$) + p($t$) & 0.060 & 1.06 & 1.71 \\
2020-09-10 & TESS & b & 90 & 207 & 120 & $o$ & 0.131 & 1.03 & 1.48 \\
2020-09-11 & TESS & c & 55 & 238 & 120 & $o$ & 0.124 & 0.98 & 1.56 \\
2020-09-12 & TESS & b & 91 & 207 & 120 & $o$ & 0.115 & 0.91 & 1.25 \\
2020-09-14 & TESS & b & 92 & 195 & 120 & $o$ & 0.120 & 0.95 & 1.34 \\
2020-09-14 & TESS & c & 56 & 216 & 120 & $o$ & 0.126 & 0.99 & 1.26 \\
2020-09-15 & TESS & d & 53 & 287 & 120 & $o$ & 0.138 & 1.09 & 1.24 \\
2020-09-16 & TESS & b & 93 & 207 & 120 & $o$ & 0.135 & 1.07 & 1.33 \\
2020-09-17 & TESS & e & 35 & 308 & 120 & $o$ & 0.129 & 1.02 & 1.03 \\
2020-09-18 & TESS & c & 57 & 240 & 120 & $o$ & 0.122 & 0.96 & 1.00 \\
2020-09-18 & TESS & b & 94 & 207 & 120 & $o$ & 0.116 & 0.91 & 1.48 \\
2020-09-18 & TESS & f & 14 & 267 & 120 & $o$ & 0.125 & 0.98 & 1.37 \\
2020-09-28 & CHEOPS & b, d, g & 99, 55, 11 & 434 & 60 & sp($r$) + p($b$) & 0.061 & 1.05 & 1.16 \\
2020-10-03 & CHEOPS & b, f & 102, 15 & 426 & 60 & sp($r$) + p($y$) & 0.057 & 1.02 & 1.25 \\
2020-10-04 & CHEOPS & d & 56 & 483 & 60 & sp($r$) + p($b$) & 0.059 & 1.05 & 2.19 \\
2020-10-05 & CHEOPS & b & 103 & 365 & 60 & sp($r$) + p($b$) & 0.057 & 1.01 & 1.31 \\
2020-10-07 & CHEOPS & c, e & 63, 37 & 486 & 60 & sp($r$) + p($y$) & 0.059 & 1.05 & 1.27 \\
2020-10-07 & CHEOPS & b & 104 & 374 & 60 & sp($r$) + $o$ & 0.056 & 1.00 & 1.58 \\
2020-10-09 & CHEOPS & b & 105 & 394 & 60 & sp($r$) + $o$ & 0.058 & 1.04 & 1.03 \\
2020-10-19 & CHEOPS & b, f & 110, 16 & 329 & 60 & sp($r$) + p($t$) & 0.057 & 1.02 & 1.27 \\
2020-10-23 & CHEOPS & c & 68 & 258 & 60 & sp($r$) + p($t$) & 0.061 & 1.08 & 1.16 \\
2020-10-28 & CHEOPS & b & 115 & 205 & 60 & sp($r$) + $o$ & 0.062 & 1.11 & 1.36 \\
2020-10-30 & CHEOPS & c & 70 & 282 & 60 & sp($r$) + p($ttt$) & 0.060 & 1.07 & 1.68 \\
2020-11-02 & CHEOPS & c & 71 & 268 & 60 & sp($r$) + p($t$) & 0.061 & 1.09 & 1.75 \\
2020-11-03 & CHEOPS & b, f & 118, 17 & 233 & 60 & sp($r$) + p($ttt$) & 0.057 & 1.00 & 1.49 \\
2020-11-06 & CHEOPS & e & 40 & 297 & 60 & sp($r$) + $o$ & 0.056 & 1.00 & 1.84 \\
2020-11-06 & CHEOPS & d & 61 & 247 & 60 & sp($r$) + $o$ & 0.059 & 1.06 & 1.90 \\
2020-11-07 & CHEOPS & b & 120 & 229 & 60 & sp($r$) + $o$ & 0.060 & 1.08 & 1.41 \\
2020-11-08 & CHEOPS & g & 13 & 223 & 60 & sp($r$) + $o$ & 0.056 & 1.00 & 2.56 \\
2021-08-04 & CHEOPS & b, f, c & 261, 35, 156 & 293 & 60 & sp($r$) + $o$ & 0.063 & 1.12 & 2.10 \\
2021-08-04 & CHEOPS & c, g & 156, 26 & 343 & 60 & sp($r$) + p($y$) & 0.066 & 1.18 & 1.48 \\
2021-10-08 & CHEOPS & b, c & 295, 176 & 365 & 60 & sp($r$) + $o$ & 0.057 & 1.01 & 2.66 \\
2021-10-15 & CHEOPS & b & 299 & 327 & 60 & sp($r$) + $o$ & 0.058 & 1.01 & 1.47 \\
2021-10-17 & CHEOPS & b & 300 & 296 & 60 & sp($r$) + $o$ & 0.067 & 1.15 & 1.43 \\
\end{longtable}

\clearpage

\subsection{Prior distributions}

\begin{table}[hbt!]
\centering
\begin{tabular}{lc}
\toprule
\toprule
Parameter (unit) & Prior \\
\midrule
\vspace{0.1cm}
Mass, $M_\star$ ($M_\odot$) & $\mathcal{N}$(0.647, 0.030$^2$) \\
\vspace{0.1cm}
Radius, $R_\star$ ($R_\odot$) & $\mathcal{N}$(0.662, 0.010$^2$) \\
\vspace{0.1cm}
Effective temperature, $T_{\mathrm{eff}}$ (K) & $\mathcal{N}$(4316, 70$^2$) \\
\vspace{0.1cm}
Quadratic limb-darkening coefficient $u_{1,\:\mathrm{CHEOPS}}$ & $\mathcal{N}$(0.538, 0.023$^2$)  \\
\vspace{0.1cm}
Quadratic limb-darkening coefficient $u_{2,\:\mathrm{CHEOPS}}$ & $\mathcal{N}$(0.157, 0.033$^2$)  \\
\vspace{0.1cm}
Quadratic limb-darkening coefficient $u_{1,\:\mathrm{TESS}}$ & $\mathcal{N}$(0.487, 0.019$^2$) \\
\vspace{0.1cm}
Quadratic limb-darkening coefficient $u_{2,\:\mathrm{TESS}}$ & $\mathcal{N}$(0.193, 0.047$^2$) \\
\vspace{0.1cm}
Quadratic limb-darkening coefficient $u_{1,\:\mathrm{NGTS}}$ & $\mathcal{N}$(0.529, 0.030$^2$) \\
\vspace{0.1cm}
Quadratic limb-darkening coefficient $u_{2,\:\mathrm{NGTS}}$ & $\mathcal{N}$(0.171, 0.045$^2$) \\
\bottomrule
\bottomrule
\end{tabular}
\caption{Prior probability distribution functions assumed in our global transit analysis (see Sect. \ref{sec:analysis}). $\mathcal{N}(\mu, \sigma^2)$ represents a normal distribution of mean $\mu$ and variance $\sigma^2$.}
\label{tab:priors}
\end{table}

\vspace{0.2cm}

\section{Results: individual transit timings}

\begin{longtable}{l c c c}
\hline
\hline
Epoch & Transit timing & TTV & Source \\
 & ($\mathrm{BJD_{TDB}} - 2\,450\,000$) & (minutes) & \\
\hline
\endfirsthead
\multicolumn{4}{c}{\tablename\ \thetable{} -- \textit{continued from previous page}} \\
\hline
Epoch & Transit timing & TTV & Source \\
 & ($\mathrm{BJD_{TDB}} - 2\,450\,000$) & (minutes) & \\
\hline
\endhead
\hline \multicolumn{4}{c}{\textit{Continued on next page}} \\ 
\endfoot
\hline \hline
\caption{Individual transit timings returned by our global transit analysis (see Sect. \ref{sec:analysis}) for the five outer TOI-178 planets. The epochs and corresponding TTVs given here are relative to the updated mean transit ephemerides given in Tables \ref{tab:transit_parameters1} (planets $b$, $c$, and $d$) and \ref{tab:transit_parameters2} (planets $e$, $f$, and $g$).}
\label{tab:timings}\\
\endlastfoot
\multicolumn{4}{c}{\textit{TOI-178\,c}} \\
\hline
\vspace{0.15cm}
-176 & $8356.0989_{-0.0138}^{+0.0080}$ & $-2.89_{-19.87}^{+11.53}$ & TESS \\
\vspace{0.15cm}
-175 & $8359.3408_{-0.0050}^{+0.0061}$ & $2.09_{-7.21}^{+8.76}$ & TESS \\
\vspace{0.15cm}
-174 & $8362.5850_{-0.0095}^{+0.0086}$ & $10.29_{-13.75}^{+12.43}$ & TESS \\
\vspace{0.15cm}
-173 & $8365.8208_{-0.0078}^{+0.0096}$ & $6.56_{-11.29}^{+13.82}$ & TESS \\
\vspace{0.15cm}
-172 & $8369.052_{-0.013}^{+0.016}$ & $-3.38_{-19.30}^{+23.47}$ & TESS \\
\vspace{0.15cm}
-171 & $8372.308_{-0.015}^{+0.011}$ & $21.06_{-22.18}^{+16.42}$ & TESS \\
\vspace{0.15cm}
-170 & $8375.5181_{-0.0052}^{+0.0077}$ & $-19.55_{-7.53}^{+11.10}$ & TESS \\
\vspace{0.15cm}
-169 & $8378.7718_{-0.0077}^{+0.0075}$ & $2.43_{-11.16}^{+10.77}$ & TESS \\
\vspace{0.15cm}
44 & $9068.5625_{-0.0086}^{+0.0104}$ & $4.00_{-12.40}^{+14.98}$ & CHEOPS \\
\vspace{0.15cm}
45 & $9071.797_{-0.012}^{+0.014}$ & $-1.78_{-17.28}^{+20.74}$ & CHEOPS \\
\vspace{0.15cm}
46 & $9075.034_{-0.010}^{+0.008}$ & $-3.44_{-14.98}^{+11.49}$ & CHEOPS \\
\vspace{0.15cm}
51 & $9091.2349_{-0.0055}^{+0.0047}$ & $8.67_{-7.99}^{+6.78}$ & TESS \\
\vspace{0.15cm}
52 & $9094.4701_{-0.0038}^{+0.0026}$ & $3.99_{-5.47}^{+3.74}$ & TESS \\
\vspace{0.15cm}
53 & $9097.7094_{-0.0094}^{+0.0082}$ & $5.20_{-13.55}^{+11.88}$ & TESS \\
\vspace{0.15cm}
55 & $9104.1840_{-0.0076}^{+0.0086}$ & $1.86_{-11.00}^{+12.36}$ & TESS \\
\vspace{0.15cm}
56 & $9107.4252_{-0.0077}^{+0.0073}$ & $5.81_{-11.10}^{+10.56}$ & TESS \\
\vspace{0.15cm}
57 & $9110.6629_{-0.0050}^{+0.0075}$ & $4.74_{-7.16}^{+10.81}$ & TESS \\
\vspace{0.15cm}
63 & $9130.0860_{-0.0018}^{+0.0031}$ & $-6.21_{-2.62}^{+4.46}$ & CHEOPS \\
\vspace{0.15cm}
68 & $9146.2850_{-0.0029}^{+0.0033}$ & $3.59_{-4.18}^{+4.77}$ & CHEOPS \\
\vspace{0.15cm}
70 & $9152.7557_{-0.0049}^{+0.0053}$ & $-5.30_{-7.07}^{+7.70}$ & CHEOPS \\
\vspace{0.15cm}
71 & $9155.9917_{-0.0097}^{+0.0089}$ & $-8.88_{-14.00}^{+12.76}$ & CHEOPS \\
\vspace{0.15cm}
156 & $9431.2758_{-0.0069}^{+0.0098}$ & $14.06_{-9.92}^{+14.05}$ & CHEOPS \\
\vspace{0.15cm}
176 & $9496.0299_{-0.0056}^{+0.0065}$ & $-7.38_{-8.02}^{+9.30}$ & CHEOPS \\
\hline
\multicolumn{4}{c}{\textit{TOI-178\,d}} \\
\hline
\vspace{0.15cm}
-61 & $8360.2385_{-0.0026}^{+0.0033}$ & $-6.00_{-3.72}^{+4.82}$ & TESS \\
\vspace{0.15cm}
-60 & $8366.8025_{-0.0021}^{+0.0020}$ & $2.97_{-3.02}^{+2.84}$ & TESS \\
\vspace{0.15cm}
-59 & $8373.3593_{-0.0082}^{+0.0040}$ & $1.58_{-11.88}^{+5.82}$ & TESS \\
\vspace{0.15cm}
-58 & $8379.9157_{-0.0028}^{+0.0033}$ & $-0.27_{-4.02}^{+4.72}$ & TESS \\
\vspace{0.15cm}
47 & $9068.4744_{-0.0029}^{+0.0039}$ & $-3.27_{-4.16}^{+5.63}$ & CHEOPS \\
\vspace{0.15cm}
48 & $9075.0314_{-0.0055}^{+0.0039}$ & $-4.22_{-7.92}^{+5.56}$ & CHEOPS \\
\vspace{0.15cm}
51 & $9094.7023_{-0.0055}^{+0.0036}$ & $-7.51_{-7.92}^{+5.13}$ & TESS \\
\vspace{0.15cm}
53 & $9107.8186_{-0.0043}^{+0.0040}$ & $-6.24_{-6.25}^{+5.75}$ & TESS \\
\vspace{0.15cm}
55 & $9120.9395_{-0.0013}^{+0.0012}$ & $1.58_{-1.83}^{+1.73}$ & CHEOPS \\
\vspace{0.15cm}
56 & $9127.4973_{-0.0022}^{+0.0022}$ & $1.74 \pm 3.17$ & CHEOPS \\
\vspace{0.15cm}
61 & $9160.2843_{-0.0018}^{+0.0016}$ & $-0.70_{-2.64}^{+2.30}$ & CHEOPS \\ 
\hline
\multicolumn{4}{c}{\textit{TOI-178\,e}} \\
\hline
\vspace{0.15cm}
-40 & $8362.9541_{-0.0031}^{+0.0058}$ & $0.07_{-4.51}^{+8.29}$ & TESS \\
\vspace{0.15cm}
-39 & $8372.9091_{-0.0091}^{+0.0085}$ & $-9.80_{-13.05}^{+12.25}$ & TESS \\
\vspace{0.15cm}
31 & $9070.2469_{-0.0039}^{+0.0041}$ & $5.65_{-5.66}^{+5.88}$ & CHEOPS \\
\vspace{0.15cm}
33 & $9090.1771_{-0.0087}^{+0.0077}$ & $15.11_{-12.60}^{+11.16}$ & TESS \\
\vspace{0.15cm}
34 & $9100.1317_{-0.0014}^{+0.0016}$ & $4.77_{-1.96}^{+2.28}$ & CHEOPS \\
\vspace{0.15cm}
35 & $9110.1036_{-0.0112}^{+0.0087}$ & $19.29_{-16.13}^{+12.54}$ & TESS \\
\vspace{0.15cm}
37 & $9130.0109_{-0.0012}^{+0.0012}$ & $-4.29_{-1.70}^{+1.71}$ & CHEOPS \\
\vspace{0.15cm}
40 & $9159.9040_{-0.0054}^{+0.0061}$ & $6.72_{-7.78}^{+8.80}$ & CHEOPS \\
\hline
\multicolumn{4}{c}{\textit{TOI-178\,f}} \\
\hline
\vspace{0.15cm}
-35 & $8364.9143_{-0.0041}^{+0.0039}$ & $-3.41_{-5.92}^{+5.57}$ & TESS \\
\vspace{0.15cm}
-34 & $8380.1582_{-0.0101}^{+0.0090}$ & $13.87_{-14.54}^{+12.99}$ & TESS \\
\vspace{0.15cm}
13 & $9096.0564_{-0.0051}^{+0.0056}$ & $8.78_{-7.39}^{+8.02}$ & TESS \\
\vspace{0.15cm}
14 & $9111.2650_{-0.0056}^{+0.0092}$ & $-24.86_{-7.99}^{+13.31}$ & TESS \\
\vspace{0.15cm}
15 & $9126.5155_{-0.0012}^{+0.0015}$ & $1.90_{-1.70}^{+2.22}$ & CHEOPS \\
\vspace{0.15cm}
16 & $9141.7445_{-0.0029}^{+0.0033}$ & $-2.36_{-4.12}^{+4.74}$ & CHEOPS \\
\vspace{0.15cm}
17 & $9156.9773_{-0.0015}^{+0.0016}$ & $-1.13_{-2.20}^{+2.25}$ & CHEOPS \\
\vspace{0.15cm}
35 & $9431.1468_{-0.0103}^{+0.0093}$ & $-9.23_{-14.83}^{+13.35}$ & CHEOPS \\
\hline
\multicolumn{4}{c}{\textit{TOI-178\,g}} \\
\hline
\vspace{0.15cm}
-26 & $8354.5502_{-0.0041}^{+0.0038}$ & $8.97_{-5.89}^{+5.46}$ & TESS \\
\vspace{0.15cm}
-25 & $8375.2603_{-0.0046}^{+0.0045}$ & $9.30_{-6.64}^{+6.51}$ & TESS \\
\vspace{0.15cm}
-6 & $8768.7409_{-0.0031}^{+0.0029}$ & $-1.81_{-4.49}^{+4.18}$ & NGTS \\
\vspace{0.15cm}
10 & $9100.0936_{-0.0025}^{+0.0041}$ & $-10.13_{-3.60}^{+5.93}$ & CHEOPS \\
\vspace{0.15cm}
11 & $9120.8067_{-0.0014}^{+0.0018}$ & $-5.61_{-2.07}^{+2.64}$ & CHEOPS \\
\vspace{0.15cm}
13 & $9162.2292_{-0.0118}^{+0.0131}$ & $-1.65_{-16.99}^{+18.86}$ & CHEOPS \\
\vspace{0.15cm}
26 & $9431.4659_{-0.0022}^{+0.0020}$ & $9.74_{-3.18}^{+2.87}$ & CHEOPS \\
\end{longtable}

\newpage

\section{Dynamical analysis: TTV mass and eccentricity posteriors for different mass priors}

\renewcommand{\arraystretch}{1.5}

\begin{table*}[hbt!]
\begin{small} 
\caption{TTV mass and eccentricity posteriors for the 5 outer planets of the TOI-178 system.} 
\label{tab:masscomp} 
\centering 
\begin{tabular}{llllllll} 
\hline 
 & Source & Prior & TOI-178\,c &  TOI-178\,d & TOI-178\,e & TOI-178\,f & TOI-178\,g\\  
\hline\hline
$M_{\mathrm{p}}$ [$M_\oplus$]& RVs (\citealt{Leleu2021})   & --  & $4.77 \pm 0.68$&$3.01 \pm 1.03$&$3.86 \pm 1.25$&$7.72 \pm 1.67$&$3.94 \pm 1.62$\\
$M_{\mathrm{p}}$ [$M_\oplus$] & TTVs & \textit{RV} & $4.70_{-0.63}^{+0.63}$&$3.35_{-0.79}^{+0.81}$&$3.11_{-0.87}^{+0.96}$&$7.50_{-1.18}^{+1.27}$&$3.12_{-1.11}^{+1.10}$\\
$M_{\mathrm{p}}$ [$M_\oplus$] & TTVs & \textit{default} & $0.22_{-0.17}^{+0.78}$&$7.73_{-1.71}^{+2.09}$&$3.61_{-1.13}^{+1.31}$&$0.78_{-0.59}^{+0.99}$&$0.17_{-0.12}^{+0.51}$\\
$M_{\mathrm{p}}$ [$M_\oplus$] & TTVs & \textit{high-mass}& $8.08_{-5.26}^{+7.98}$&$8.33_{-2.23}^{+2.33}$&$3.12_{-1.45}^{+1.64}$&$6.76_{-2.51}^{+2.62}$&$1.88_{-1.19}^{+1.47}$\\
$\Delta_M$ & Robustness criterion & -- & $10.08$&$0.29$&$0.37$&$6.06$&$3.34$ \\
$e_{\mathrm{p}}$ & TTVs & \textit{RV} & $0.0073_{-0.0051}^{+0.0083}$&$0.010_{-0.007}^{+0.011}$&$0.0080_{-0.0057}^{+0.0100}$&$0.0105_{-0.0061}^{+0.0071}$&$0.0056_{-0.0039}^{+0.0058}$\\
$e_{\mathrm{p}}$ & TTVs & \textit{default} & $0.0071_{-0.0050}^{+0.0081}$&$0.0051_{-0.0036}^{+0.0067}$&$0.0041_{-0.0029}^{+0.0054}$&$0.012_{-0.008}^{+0.016}$&$0.022_{-0.016}^{+0.027}$\\
$e_{\mathrm{p}}$ & TTVs & \textit{high-mass} & $0.0033_{-0.0023}^{+0.0038}$&$0.0049_{-0.0034}^{+0.0055}$&$0.0050_{-0.0035}^{+0.0057}$&$0.0068_{-0.0045}^{+0.0065}$&$0.0039_{-0.0028}^{+0.0045}$\\
\hline 
\end{tabular} 
\end{small} 
\end{table*}

\end{appendix}

\end{document}